\documentclass[aps,pre,reprint,superscriptaddress]{revtex4-1}

\usepackage{amsmath}
\usepackage{hyperref}
\usepackage{graphicx}
\usepackage{xcolor}
\usepackage[normalem]{ulem}

\usepackage{tikz} 

\graphicspath{{data/}}

\begin{document}

\title{Ground-state properties of the two-dimensional SWAP
  spin-glass ensemble}

\author{Alexander K. Hartmann}
\affiliation{Institute of Physics, University of Oldenburg, Oldenburg, Germany}

\author{Leticia F. Cugliandolo}
\affiliation{Sorbonne Universit\'e, Laboratoire de Physique Th\'eorique et Hautes Energies, 
CNRS UMR 7589, 4 Place Jussieu, 75252 Paris Cedex 05, Paris, France}

\author{Marco Tarzia}
\affiliation{Sorbonne Universit\'e, Laboratoire de Physique Th\'eorique de la Mati\`ere Condens\'ee, 
CNRS UMR 7600, 4 Place Jussieu, 75252 Paris Cedex 05, Paris, France}

\date{\today}

\begin{abstract}
  We study the recently introduced SWAP ensemble for Ising spin-glasses,
  where the spins obtain varying lengths, with length scale $\Delta \in [0, 2]$.
  The lengths can be exchanged
  in the spirit of the SWAP algorithm for glassy poly-disperse
  hard-sphere systems.
  Using an annealing schedule, if the annealing is slow enough,
  ground states of the corresponding Ising Hamiltonian, where the spin lengths are incorporated into the bonds, can be obtained
  with high probability. We prove this statement for two-dimensional
  systems with  sizes of up to $L^2=1024^2$
  spins by comparison with exact ground states obtained by using graph matching algorithms.
  In particular, we analyze the nature of the obtained realizations
  of the disorder by calculating exact zero-temperature
  domain-wall energies, from exact ground state calculations
  using  periodic and anti-periodic boundary conditions.
  For medium or large values of $\Delta$, e.g. $\Delta=1$, the realizations
  turn ferromagnetic if the annealing is slow enough, i.e., they lose
  their zero-temperature spin-glass property. If $\Delta$ is small,
  the realization remain glassy, but the SWAP annealing does not easily find a true ground state.
\end{abstract}

\maketitle

\section{Introduction}

Spin glasses~\cite{binder1986,mezard1987,mydosh1993,young1998,nishimori2001,charbonneau2023} are 
physical examples of systems with competing interactions that lead to complex behavior. They find realizations and applications in diverse fields, such as disordered magnets, neural networks, optimization problems, the benchmarking of quantum algorithms, and economics. There is even a spin-glass board game available~\cite{spin-glass-game2013}.

Treating spin-glasses numerically is notoriously difficult, so much so that even dedicated computers such as JANUS I and JANUS II~\cite{janus} have
been developed. In contrast, for another complex physical
system--namely, poly-disperse structural glasses~\cite{binder2011,berthier2011}--
fast Monte Carlo algorithms based on swapping particle
sizes non-locally~\cite{berthier2016,ninarello2017,berthier_efficient_2019,Scalliet22},
lead to a dramatic acceleration of simulations.
Inspired by this success, the SWAP ensemble was recently introduced for
Ising spin glasses, which were
turned artificially into poly-disperse systems
by introducing uniaxial spins of varying lengths~\cite{miranda2024}.
Thus, a SWAP algorithm was developed in the spirit of
the corresponding algorithms for structural glasses.
Using a mean-field calculation, it was shown that at least some 
of the spin-glass properties of the configurations obtained via the SWAP algorithm are preserved.
Moreover,  the SWAP approach was able
to find true ground states for two-dimensional systems of 
moderate size ($N = 32^2 $ spins).

Here, we extend the analysis of the SWAP ensemble 
to substantially larger system sizes. This is 
possible in two dimensions because we use minimum-weight perfect
matching algorithms on graphs to obtain the exact ground states 
in polynomial time.
Thus, we address systems up to $N=1024^2$ spins.
Furthermore, combining this approach with a change in boundary conditions allows us 
to investigate whether the system has a stable ordered phase at finite temperatures, 
and to determine whether this phase exhibits ferromagnetic or spin-glass order.

The paper is organized as follows. First, we present the SWAP model
and some of the quantities we measure in Sec.~\ref{sec:model}.
Next, in Sec.~\ref{sec:algs} we explain the two simulation approaches that we have used.
In the main section~\ref{sec:results} we show our results for medium and small values
of the SWAP length scale, respectively. In particular, we measure the probability
that the SWAP algorithm finds true ground states, the annealing time needed to achieve this goal,
the total satisfaction (to be defined in Sec.~\ref{sec:model}) 
in the system, and the distribution of domain-wall
energies. We conclude with a summary and discussion in Sec.~\ref{sec:discussion}.

\section{Models\label{sec:model}}

We consider a model \cite{miranda2024} of spin-glasses on finite-dimensional
lattices consisting of  $N$
spins $s=(s_1,\ldots,s_N)$
with
 the Hamiltonian
\begin{equation}
{\cal H}_J(s) = \sum_{\langle ij \rangle} J_{ij} s_i s_j  \,,
\label{eq:H}
\end{equation}
where the sum  $\langle i,j \rangle$ runs over pairs of  neighboring
sites $i$ and $j$ on the lattice. We focus on planar square geometry. 
Aside from the fact that working in two dimensions is less 
demanding numerically, a renewed interest in quasi-two-dimensional cases has been recently triggered by experiments and extensive simulations~\cite{fernandez_out--equilibrium_2018, fernandez_dimensional_2019, fernandez_experiment-oriented_2019,ZhaiOrbach24,Pekarek25,Dahlberg25,li2026dynamicsspinglassesdimensions}
as well as by relations to quantum computation problems~\cite{Preskill02,Wang03,Katzgraber09,Kubica18,Agrawal23}.
The interactions $J_{ij}$ are quenched random
bimodal variables where $J_{ij}=1$ and $J_{ij}=-1$ appear both with
probability $1/2$. Realizations of these bonds correspond to the
standard Edwards-Anderson (EA)~\cite{edwards1975} model with
$\pm J$ bonds, with coupling strength usually set to one.
The spin variables can be written as 
\begin{equation}
s_i = \sigma_i \tau_i
\; ,
\end{equation}
where $\sigma_i=\pm 1$ are their signs   and $\tau_i>0$ are their lengths.
Therefore, the Hamiltonian depends on these two variables,  
${\cal H}(s)\equiv {\cal H}(\sigma,\tau)$ with $\tau=(\tau_1,\ldots,\tau_N)$ and $\sigma=(\sigma_1,\ldots,\sigma_N)$.
The length variables are contained in an interval of width $\Delta$ around
1, i.e.,
\begin{equation}
  \tau_i \in [1-\Delta/2, 1+\Delta/2]\,,
\end{equation}
where the length scale $\Delta\in[0,2]$ describes the scatter of the spin
lengths.  The lengths are assigned randomly 
at the beginning of each simulation with uniform probability within
this interval. During the simulation, we will allow the bond lengths
to fluctuate, but only through swaps between pairs of spins. Consequently, the
total distribution of length values within a realization does not change. In this sense, the $\tau$ variables are not full degrees of freedom. 
Since we are interested in the low-energy behavior of this model, this constraint prevents groups of spins from simultaneously shrinking or expanding to easily minimize the total energy. This is also inspired by the mentioned SWAP methods applied to 
glassy models, where the 
ensemble of particle sizes is fixed.
For $\Delta=0$ all lengths are $\tau_i=1$ and
the original EA model~\cite{edwards1975}
is recovered with Ising spins $s_i=\sigma_i$ for all $i$.

One can also interpret the system with varying spin lengths as a proper Ising system,
if one uses the spin length to create
 ``annealed'' bonds, which are therefore given by
\begin{equation}
{\cal J}_{ij} = J_{ij} \tau_i \tau_j 
\label{eq:cal:J}\,,
\end{equation}
and the Hamiltonian becomes
\begin{equation}
  {\cal H}_{\mathcal J}(\sigma)=\sum_{\langle ij\rangle} {\cal J}_{ij} \sigma_i \sigma_j\,.
\label{eq:H:annealed}
\end{equation}
For $\Delta$ uniformly distributed  in  $[0,2]$, the distribution of the now continuous couplings ${\mathcal J}_{ij}$ 
is symmetric around zero and has 
variance 
\begin{equation}
\label{eq:var:J:delta}
\sigma^2_{\Delta} =    J^2 (1 + \Delta^2/12)^2\,.
\end{equation}
For $\Delta \to 0$ the distribution 
is just two delta peaks at $\pm J$, with 
$\sigma^2_{\Delta}=J^2$. For 
increasing  $\Delta$
 the gap between the peaks' positions shrinks until closing 
 completely and leaving a single peak at 0 for
$\Delta = 2$, now $\sigma^2_{\Delta}= 16/9 \, J^2$, see Fig.~SM16 in \cite{miranda2024}. This means that the magnitude of the couplings does not 
change much for varying $\Delta$ in the interval $[0,2$], 
it is always order $O(J)$.

We will perform thermal annealing dynamics, see Sec.~\ref{sec:algs}, where the $\tau$ and $\sigma$ variables change with simulation time $t$, which makes these annealed bonds ${\cal J}_{ij}\equiv {\cal J}_{ij}(t)$ a function of time $t$ as well. At the final annealing time $t_f$, we fix the  
bonds to the values reached and denote them as 
\begin{equation}
{\cal J}_{ij}^*={\cal J}_{ij}(t_f)\,.
\end{equation}

The ground-state energy of Hamiltonian Eq.~(\ref{eq:H:annealed}) with fixed length variables $\tau$, i.e., fixed annealed bonds 
${\cal J}^*$, is given by
\begin{equation}
E_0 = \min_{\sigma} {\cal H}_{{\mathcal J}^*}(\sigma) \,,
\end{equation}
where the minimization is performed over all possible $2^N$ assignments of the
Ising spins $\sigma=(\sigma_1,\ldots,\sigma_N)$. Note that  the Hamiltonian  Eq.~(\ref{eq:H}) also exhibits a ground state (GS) 
which is defined with respect to both $\sigma$ and constrained $\tau$ degrees of freedom, but this, as in the original work~\cite{miranda2024}, is not explicitly considered here because the main driving interest is still the Ising spin glass with $\sigma$ degrees of freedom only. Still, as we explain below, what we learn about the physical nature of the GS of Hamiltonian Eq.~(\ref{eq:H:annealed}) tells us also the nature of the  
GSs of Hamiltonian Eq.~(\ref{eq:H}).

One main ingredient creating the complex behavior of spin-glasses is the presence of \emph{frustration}~\cite{toulouse1977,vannimenus1977,diep2025}. 
Frustration means that not all bonds of a realization can be simultaneously satisfied, that is, get  negative contributions to the energy
from all bonds. 
To quantify this feature, for each plaquette
$P=\{ {\cal J}_{ij},\,{\cal J}_{jk},\,{\cal J}_{kl},\,{\cal J}_{li} \}$
      of four neighboring bonds, and correspondingly for plaquettes of ${\cal J}_{ij}^*$ bonds,
we define the local frustration as an extension of the
``frustration function''~\cite{vannimenus1977} by also considering the magnitudes of the bonds:
\begin{equation}
  \label{eq:f:P}
f_{P} = \prod_{(i,j) \in {P}} {\cal J}_{ij}\,.
\end{equation}
For $f_P<0$, the plaquette is frustrated, i.e., it exhibits
an odd number of negative bonds. This means that it is impossible
to satisfy all bonds simultaneously for such a single plaquette.
For $f_P>0$, the plaquette is
not frustrated,  and all bonds can be simultaneously satisfied, which happens,
e.g., in all plaquettes of the ground state of a ferromagnet.

Since a positive value of $f_P$ denotes the absence of frustration, which is a bit counter-intuitive, we use a different name and define
the total 
satisfaction $s_{\rm tot}$
of an ensemble by first averaging $f_{P}$
over all plaquettes ${P}$ of a realization,
and then averaging the result over all corresponding bond realizations:
\begin{equation}
  s_{\rm tot} = \left[ \left\langle f_{P}
    \right\rangle_{P} \right]\, .
  \label{eq:stot:def}
\end{equation}
Here, $\langle \ldots\rangle_{P}$ denotes the average over all plaquettes
and $[\ldots ]$ indicates the average over all bond realizations.  Now, the larger  
$s_{\rm tot}$, the less frustrated the system is.
 We define 
two ensembles of bonds, where we measure $s_{\rm tot}$, in the next Section.

\section{Algorithms\label{sec:algs}}

The aim of this study is to generate low-energy configurations
of the realizations under investigation. We consider two
types of bond realizations: First, realizations of bonds 
$J=\{J_{ij}\}$, where the degrees of freedom are the $\tau$ and 
$\sigma$ variables. Second, realizations of bonds ${\cal J}^*=\{{\cal J}^*_{ij}\}$, 
which are obtained after annealing runs and then kept fixed for further treatment and analysis, 
such that only the $\sigma$ variables remain as degrees of freedom.

We use two types of approaches. First, we utilize algorithms mimicking physical
local and non-local dynamics, in the form of simulated annealing, as 
explained in Sec.~\ref{sec:simulated-annealing}. These algorithms are applied both to the $J$ realizations, 
to obtain ${\cal J}^*$ realizations by changing $\tau$ and $\sigma$ degrees of freedom, and to the ${\cal J}^*$ 
realizations, but only with $\sigma$ degrees of freedom.

Secondly, we use minimum-weight perfect matching
algorithms, which allow us  to obtain exact
ground state configurations of planar Ising spin-glasses, as explained in Sec.~\ref{sec:matching}. 
This approach can only be applied to systems with Ising $\sigma$  degrees of freedom, i.e., 
here mainly to systems with the ${\cal J}^*$ realizations
of bonds, and in a few cases also to systems with the $J$ realizations
of bonds, 
but with all equal lengths, $\tau\equiv 1$.

Finally, in Sec.~\ref{sec:domain-walls} we explain how we create domain walls in the ground states, measure its energy $\Delta E_0$, determine  whether a system exhibits a stable low-temperature
phase and what is the nature of the phase from the distribution of
$\Delta E_0$. This is also applied to the ${\cal J}^*$ realizations, and 
the obtained nature of the phase is either a $T=0$ spin-glass or a ferromagnet.

\subsection{Simulated Annealing \label{sec:simulated-annealing}}

The basic idea of simulated annealing \cite{kirkpatrick1983} is
to mimic in simulations how ground states are found by slowly
cooling a physical probe. Thus, one simulates the system in the
canonical ensemble, i.e., describing
a coupling to a heat bath at temperature $T$,
and slowly decreases the temperature.

We follow the procedure introduced in~\cite{miranda2024}.
For simulating the canonical ensemble of Hamiltonian Eq.~(\ref{eq:H}), we apply the Metropolis algorithm~\cite{newman1999}.
This means that we create a Markov chain of configurations $(\sigma(k),\tau(k))$, with 
$k=0,1,2,\ldots$. At any step $k$,  with current configuration
$(\sigma(k),\tau(k))$, a \emph{trial configuration} $(\sigma',\tau')$ is
generated as follows:
First, one copies the current configuration to the trial one.
Then a \emph{move} is performed.
We consider two types of moves, where the parameter $p_{\rm swap}$
controls the share between the two moves. 
\begin{itemize}
\item
  With probability
$p_{\rm swap}$ two sites $i$ and $j$ are chosen out of all sites in the sample, 
  each with uniform probability $1/N$, but with the constraint that they are 
  neither the same nor first neighbours on the lattice,
so as to make the move non-local. Then, the length values and the signs at the chosen sites
are exchanged, i.e. $\tau'_i \leftrightarrow \tau'_j$ and
$\sigma'_i \leftrightarrow \sigma'_j$.

\item Alternatively, i.e., with probability $1-p_{\rm swap}$, a single spin site
  $i$ is selected randomly again with uniform probability $1/N$ and the spin
  sign is flipped, i.e. $\sigma'_i \to -\sigma'_i$.
\end{itemize}
We use $p_{\rm swap}=0.2$
in this work.

The trial configuration is accepted, i.e.,
$(\sigma(k+1),\tau(k+1))=(\sigma',\tau')$ with the Metropolis
acceptance probability
\begin{equation}
  p_{\rm acc}=\min \{1, \exp(-\Delta {\mathcal H}/T)  \}  \,,
\end{equation}
where $\Delta {\mathcal H} = {\mathcal H}(\sigma',\tau')- {\mathcal H}(\sigma(k),\tau(k))$
is the energy difference between trial and current configuration.
Else, with probability $1-p_{\rm acc}$, the current configuration
  at step $k$ remains
  the configuration at step $k+1$.
  To measure time $t$ we use \emph{sweeps} which
  consist of $N$ steps, i.e., $t=k/N$.
  At time  $t=0$, we start all SWAP annealing
  runs with an initial temperature $T=T_0$,
  with a random assignment of the spin signs $\sigma_i=\pm 1$,
  and with the lengths randomly chosen with uniform probability in the
  interval of length $\Delta$ around the value 1.

We also use the Metropolis algorithm for the case where only the $\sigma$ variables are degrees of freedom, 
in particular when we consider the ${\cal J}^*$ realizations within Hamiltonian Eq.~(\ref{eq:H:annealed}). 
This means that we perform only local spin flips which is achieved by setting  $p_{\rm swap}=0$.
  
  To perform annealing, during the simulation the temperature is gradually
  decreased. For the temperature schedule $T(t)$ as a function of the
  simulation time $t$, we apply a quadratic cooling schedule
\begin{equation}
  T(t)=T_0 \, (1-t/t_f)^2\;,
  \label{eq:annealing}
\end{equation}
where 
$T_0$ is the initial temperature and the final time $t_f$ denotes the total
number of Monte Carlo sweeps. This is the same cooling protocol
used in~\cite{miranda2024}. Since two-dimensional spin-glasses
do not exhibit a finite-temperature phase transition, any value $T_0>0$
is located in the paramagnetic phase and should therefore
yield similar results. We have selected $T_0=1$ throughout this 
work.

For simulated annealing,
exponential cooling  is often applied, i.e. $T(t)=T_0 \, f_T^t$ where
$f_T$ is a factor smaller than but close to one, e.g., $f_T=0.99$. It has been shown 
that using this schedule with  infinitesimal small cooling, $f_T\to 1$, 
a ground state is reached with probability one~\cite{german1984}. 
Clearly, for practical simulations the available
time is finite and this procedure becomes rapidly impracticable. 

We preferred to use the power-law cooling in Eq.~(\ref{eq:annealing}) 
for two reasons: it was successfully applied 
in the annealing of two-dimensional
spin-glasses~\cite{rubin2017} and it was used in
the publication~\cite{miranda2024} where the SWAP ensemble was introduced,
making the past and present results comparable. Note also that the
quadratic schedule stays longer at higher temperatures allowing for
stronger fluctuations, which works well for finite systems and
finite annealing times, as compared to the exponential cooling, which
is proven to lead to ground states only in the infinite slow case
$f_T\to 1$.

The Ising spin configurations $\sigma(t_f)$ reached at the end of the annealing we denote
by $\sigma^*$, corresponding to the final annealed bond realizations denoted by ${\cal J}^*$.

\subsection{Exact Ground States \label{sec:matching}}

The calculation of exact GS $\sigma^0$ for the Hamiltonian in 
Eq.~(\ref{eq:H:annealed}) by using graph matching algorithms
can be applied to planar spin-glasses.  Whenever we speak of 
GS in this work, we always refer to the 
minimization of Eq.~(\ref{eq:H:annealed}), except explicitly stated otherwise. 

Mainly, we apply the approach to realizations ${\cal J}={\cal J}^*$ of bonds stored at the end of the annealing process.
Still, sometimes we also apply it to the original $\pm J$ realizations with ${\cal J}=J$, fixing 
$\Delta=0$ since then all $\tau_i=1$ and we recover Ising spins.

Here, we outline only the 
approach~\cite{thomas2007,pardella2008} that we used, which is also
explained in detail in the literature~\cite{khoshbakht2018}.
For readers interested in the details needed to actually implement
the algorithm, we
refer to the cited literature.

The approach is based on the transformation of a realization into an auxiliary graph $K$, where the edges of the graph carry weights.
For this purpose, each plaquette $P$
of the realization  is represented in $K$ by a set of four nodes,
called \emph{Kasteleyn cities}. Within a city,
all four nodes are
connected to each other by edges with zero weight. For the
bonds ${\cal J}$ of a realization, additional edges with weights ${\cal J}_{ij}$  are added in $K$, connecting the cities that represent the
plaquettes which are shared by the bonds ${\cal J}_{ij}$, respectively.
 Also, the free
boundary conditions have to be taken into account, by adding more cities and edges; for
the actual details, see the literature \cite{thomas2007,pardella2008,khoshbakht2018}.

Now, one can show \cite{thomas2007,pardella2008}
that when calculating a \emph{minimum-weight perfect matching} (MWPM)
for $K$, one obtains a set of loops all exhibiting a negative weight,
where each loop surrounds spins of
the same orientation in the ground state. Thus, from this set of loops a ground state
spin configuration $\sigma^0$ can be obtained. For the actual calculation of the MWPM
we use an implementation from the LEDA library \cite{leda1999}.
The calculation of a MWPM for graphs with $\tilde N$ nodes
and $\tilde M$ edges
exhibits a worst-case time complexity of $O(\tilde N \tilde M \log \tilde N)$. Here, for the Kasteleyn graph $K$, we have 
$\tilde N\approx 4N$ and $\tilde M \approx 8N$.
The approach which is explained here for square
lattices can also be used for other lattice geometries. For instance, it
has been applied to study Ising spin-glasses and random-bond models
on triangular lattices~\cite{ccp2011}.

Importantly enough,  there is another  approach \cite{bieche1980} to obtain spin-glass
ground states of planar spin-glasses in polynomial time.
It is based on a different
construction of auxiliary graphs. It has also been
widely applied~\cite{stiff2d,aspect-ratio2002,droplets2003,rand_bond2004,droplets_long2004} for various related studies.
 For this approach the size of the equivalent
graphs grows faster in the size of the original system
as compared to the approach based on Kasteleyn cities. Thus, the
Kasteleyn cities more recent approach
allows one to treat larger systems and is therefore
applied in this work.

\subsection{Domain Walls\label{sec:domain-walls}}

A powerful approach used to investigate whether a system exhibits an ordered phase 
is to verify its stability against large-scale, low-energy fluctuations~\cite{bray1984}.
To this end, we consider system-spanning domain-walls, which are
introduced by flipping the boundary conditions in one direction.
More precisely, we first calculate the ground state (GS) of the given
disorder realization with \emph{periodic}
boundary conditions in one direction, say the $x$ direction. The  GS energy
is denoted as $E_0^{\rm p}$.
Then, we switch the boundary conditions in this direction
from periodic to \emph{anti-periodic} by changing the sign of the bonds
along one column, i.e., wrapping around in the $x$ direction.
Next, we calculate the new GS, yielding the GS energy $E_0^{\rm ap}$.

The domain wall energy is  the difference of the GS energies of the
system with  periodic and anti-periodic boundary conditions, 
\begin{equation}
\label{eq:de0}
\Delta E_0 = E_0^{\rm p}-E_{0}^{\rm ap}\,.
\end{equation}
In general, the ground states could be degenerate. Nevertheless, here we focus  only on their energies, 
so the degeneracy is not important. Moreover, since the realizations ${\cal J}$ are obtained by multiplying $\pm J$ 
bonds with real numbers $\tau_i$,  the ground states of Hamiltonian Eq.~(\ref{eq:H:annealed}) should be unique.

In a generic case, the difference $\Delta E_0$ may be positive or negative. 
For a ferromagnetic system with $N=L^d$
spins, $\Delta E_0$ is  positive and grows like
$\Delta E_0 \sim L^{d-1}$.
For a random ferromagnetic model the corresponding average
$[ \Delta E_0 ]$ over the disorder realizations
also shows this scaling behavior. The growth
 with a positive exponent for $d>1$ signifies the existence of a
 stable ferromagnetic phase
at low temperatures.

For a spin-glass, on the other hand, due to the randomness and frustration, 
$\Delta E_0$ fluctuates around 0, i.e. $[ \Delta E_0 ] \to 0$
in the limit of infinite system sizes.
Note that the described domain-wall calculation
is essentially equivalent to performing
a block-spin transformation with a block-spin size of $L^d$  within real-space normalization. 
Thus, $\Delta E_0$
is an effective coupling constant of the block-spin with its neighbors.
For a spin-glass the distribution of bonds is initially random and will
be random after a block-spin transformation as well but, although modified, the
symmetry constrains it to remain centered at zero.

The width 
\begin{equation}
  \sigma_{\Delta E_0} = \sqrt{[ \Delta E_0^2 ]
    - [  \Delta E_0 ]^2}
\label{eq:def:width}
\end{equation}
of the
 block spin or domain-wall energy distribution, i.e., the typical strength, is crucial.
This is so because in the canonical ensemble the ratio $J/T$ of bond strengths and the temperature enters the Boltzmann weight.
The width often behaves like
a power law
\begin{equation}
  \sigma_{\Delta E_0}\sim L^{\theta}\,.
\label{eq:def:theta}
\end{equation}
The renormalization view  tells us whether a system
exhibits a stable spin-glass phase. This is the case if the distribution
widens, i.e., $\theta>0$, because typical bonds, let them be positive or
negative, between block spins gain strength, which enforces any order.
On the other hand if $\theta<0$, the spin-glass will not have
a stable finite-temperature phase.
In this way, it could be shown by exact GS domain-wall calculations~\cite{stiff2d,aspect-ratio2002}
that the two-dimensional spin-glass
does not exhibit a stable low temperature phase. For
Gaussian randomness a value $\theta$ about $-0.28$ was found,
where the highest-precision value was obtained
as $\theta=-0.2793(3)$~\cite{khoshbakht2018}.
For dimensions larger than two, positive exponents where
reported~\cite{stiff3d,stiff3d,boettcher2004b,boettcher2005}
indicating a stable spin-glass phase, but here heuristic
algorithms were applied to  obtain GSs due to
the lack of polynomial-time exact GS approaches.

Other methods have been developed to study random systems
based on changing bonds and then recalculating the GSs.
For instance, by systematically introducing selected sets of
\emph{hard} bonds, i.e., with very large magnitudes such that they always
have to be satisfied, droplet excitations were obtained
\cite{droplets2003,droplets_long2004}. This yielded
a numerical proof that the droplet theory~\cite{mcmillan1984,bray1986,fisher1986,fisher1988} 
is valid for
two-dimensional spin-glasses.

\section{Results\label{sec:results}}

In this section we present our  numerical study \cite{practical_guide2015} 
of two-dimensional spin glasses for system sizes between $L=8$ and $L=1024$. 
For all simulations we performed an average over 1000 realizations of the 
original disorder $\{J_{ij}\}$.

\subsection{Ground-State Energy}

\begin{figure}[ht]
\vspace{0.5cm}
    \includegraphics[width=0.9\columnwidth]{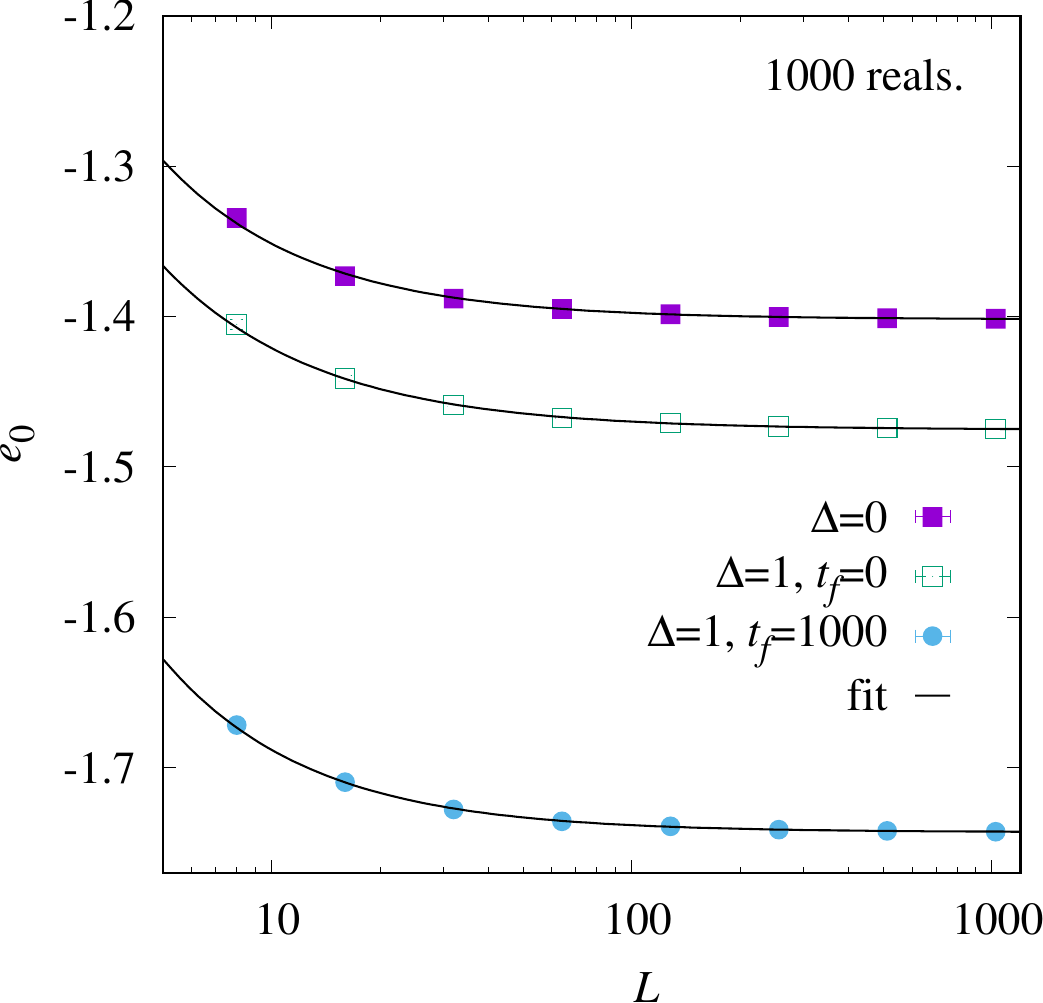}
    \caption{\label{fig:e0_L}
      The GS energy density $e_0$ of the model with Hamiltonian in Eq.~(\ref{eq:H:annealed}) as a function of the system size $L$.
      Shown are three types of disorder realizations ${\cal J} = \{ {\cal J}_{ij} \}$: the
      original systems ${\cal J}_{ij}=J_{ij}$, i.e. $\Delta=0$, with filled violet data points,
      the modified bonds ${\cal J}_{ij} (t_f=0)$ for
      $\Delta=1$ without annealing with open green data points,
      and the
    case of the bonds ${\cal J}_{ij}^*$ for
    $\Delta=1$ reached after $t_f=10^3$ SWAP annealing sweeps with filled cyan datapoints. Each time an average over 1000 realizations was performed. The solid lines represent fits
    to the function in Eq.~(\ref{eq:e0:fit}), respectively.
        }
\end{figure}

We first study the disorder average of the exact
GS energy densities  $e_0=[ E_0 ] /N$ for 
different types of bond realizations ${\cal J} = \{ {\mathcal J}_{ij} \}$,
 but always minimizing the Ising Hamiltonian Eq.~(\ref{eq:H:annealed}). We show $e_0$ as a function
of system size in Fig.~\ref{fig:e0_L}. All GSs are obtained
by using the exact matching approach explained in Sec.~\ref{sec:matching}.

For the original $\pm J$ spin-glass case,
which corresponds to 
$\Delta=0$ and therefore ${\cal J}\equiv J$, we find energy densities above and near $e_0=-1.4$. 
To extrapolate to the infinite system size, we fit a power law plus a
constant of the form
\begin{equation}
  \label{eq:e0:fit}
  e_0(L) = e_0^{\infty}+bL^{-c}\,,
\end{equation}
where $e_0^{\infty}$ is the extrapolated GS energy density
and $b$ and $c$
describe the finite-size behavior. 

Firstly, for the case $\Delta=0$, we obtain
$e_0=-1.40186(5)$,
which is compatible with the past 
estimate $e_0=-1.40193(2)$ which was 
measured with large-scale exact GS calculations~\cite{palmer1999}.

Secondly, we consider a SWAP system with $\Delta=1$, i.e. varying lengths
and therefore annealed bonds  having not only $\pm 1$ values, but  without actually performing an annealing, i.e., the bonds ${\cal J}(t_{f}=0)$. 
For these modified bonds we compute their exact GSs.
As visible in the middle curve of Fig.~\ref{fig:e0_L} (open green datapoints),
this allows for lower GS energies. From the fit we obtain
$e_0=-1.47526(6)$.
The lower GS as compared to the original $\pm J$ model
can be understood as follows: the original system is highly degenerate,
there is an exponential number of GSs, all exhibiting the same
energy. By introducing the scaled SWAP bonds ${\cal  J}(t_f=0)$,  the degeneracy is
lifted. Therefore, some of the previous GSs attain higher
and some lower energies, and the lowest one is selected, maybe even a
configurations which was not a GS for the original $\pm J$ realization.
Thus, the GS of the system with the scaled bonds
will be at lower energy. 
Note that the ratio of ground state energies is $e_0(\Delta =1)/e_0(\Delta=0) =1.05$. This is of the order of the change of the magnitude of the bonds as given by  Eq.~(\ref{eq:var:J:delta}) which is
here $\sqrt{\sigma^2_{\Delta=1}}/\sqrt{\sigma^2_{\Delta=0}}=13/12\approx 1.08$.

Thirdly, we study the annealing case.
When reaching $T=0$ after $t_f=1000$ SWAP annealing
sweeps, for the  obtained configuration ${\cal J}^*$, an exact GSs is calculated.
Since SWAP annealing aims at minimizing Hamiltonian Eq.~(\ref{eq:H}), it can be 
expected that the such obtained annealed realizations ${\cal J}^*$ also exhibit lower GS energies    
with respect to Hamiltonian Eq.~(\ref{eq:H:annealed}). 
This is visible in the lower part of
Fig.~\ref{fig:e0_L} (filled cyan datapoints), 
and an extrapolated value of $e_0=-1.74286(8)$
is obtained. 
Such a low GS energy density is already rather close
to the one of the pure two-dimensional ferromagnet which is $e_0=-2$. 
Since the standard deviation 
$\sqrt{\sigma^2_{\Delta=1}}$
is $\sim 1.08 \, J$, the energy scales in the disordered and FM models are
approximately the same. Thus, the lower GS energy found for the 
models ${\mathcal J}^*$ should be due to correlated rearrangements of the spin lengths and the corresponding changes of the couplings, and not only originate from changes in the magnitudes of the couplings.
Thus, the question arises as to whether the SWAP annealed Ising realizations   ${\mathcal J}^*$
bear similarity with ferromagnets, a question that we will
answer in Sec.~\ref{sec:domain_wall}. 

Before addressing the issue stressed at the end of the previous paragraph, 
we investigate whether simulated annealing actually finds the GSs and, if so, how long 
does the process take. We do this because the application of SWAP moves was originally 
motivated by the search for very low-lying configurations—including the GSs—more efficiently than by using only physical single-spin flips.

\subsection{Ground-State Probability \label{sec:gs_probability}}

The exact GSs $\sigma^0$
obtained by the matching algorithm, as outlined in
Sec.~\ref{sec:matching}. For the realizations
with modified bonds  can be compared with the
spin configurations $\sigma^*$ obtained at the end of the annealing, for the same
set of final bonds ${\mathcal J}^*$. 
If the SWAP annealing works well, confirming the initial motivation
for this ensemble, it should yield the ground state energy of 
the Hamiltonian in Eq.~(\ref{eq:H:annealed}) 
as the exact GSs, i.e. ${\mathcal H}_{{\mathcal J}^*}(\sigma^0) = {\mathcal H}_{{\mathcal J}^*}(\sigma)$.

\begin{figure}[ht]
    \includegraphics[width=0.9\columnwidth]{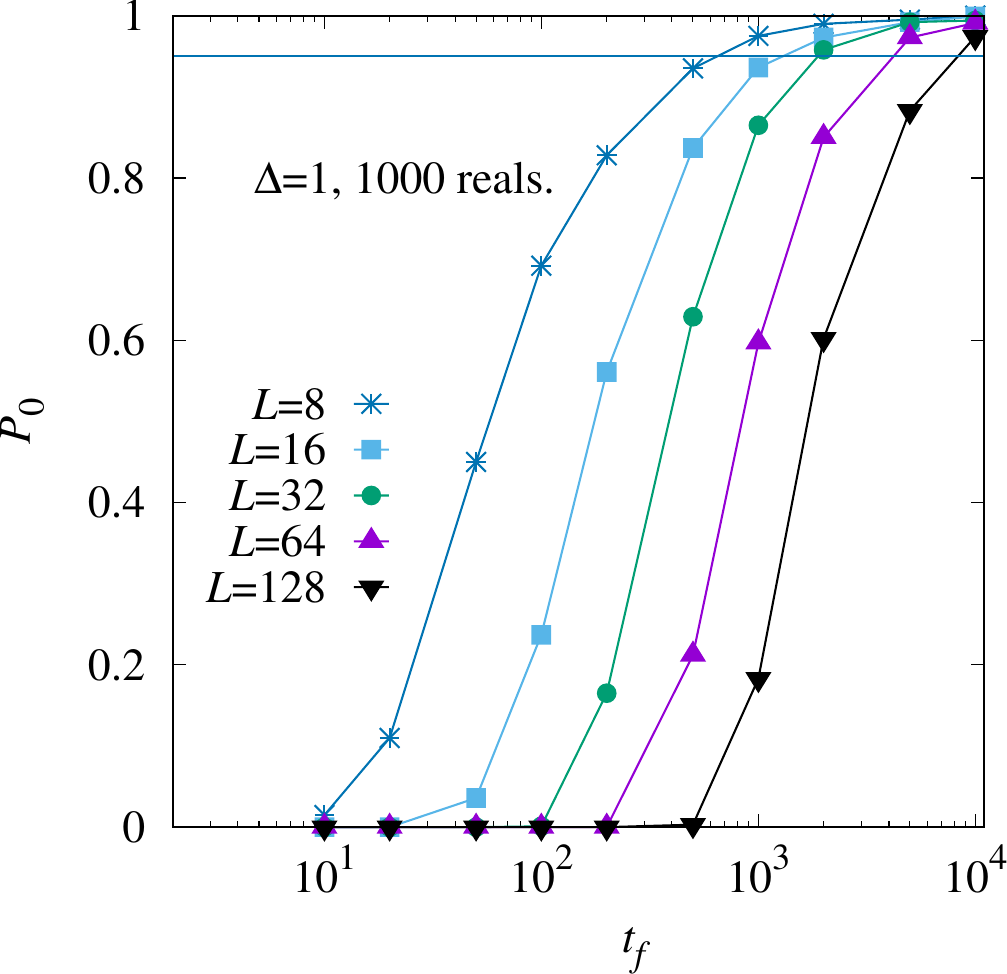}
    \caption{\label{fig:p0_tmc_delta10}
      The probability $P_{0}$ that at the end of the square annealing 
      procedure in Eq.~(\ref{eq:annealing}) a GS
      is obtained for the resulting disorder realization ${\mathcal J}^*$
      as a function of the number $t_f$ of annealing sweeps, for
      $\Delta=1$ and  different system sizes $L$. The lines joining the data points are guides to the eyes only. 
      The horizontal blue line represents the $P_0=0.95$ level. 
      }
\end{figure}

In Fig.~\ref{fig:p0_tmc_delta10} we show the probability $P_0$
that the configuration $\sigma^*$ obtained at the end of the annealing
is an exact GS. 
Here, $P_0$ is estimated
from 1000 realizations of the initial disorder $\{ J_{ij} \}$, respectively.

With growing annealing time $t_f$ the
probability to find a GS increases monotonically and indeed reaches
one for slow-enough annealing. Thus, the original motivation that 
GSs can be found by the SWAP annealing is satisfied.

In the original publication~\cite{miranda2024} the question
as to whether GSs are found was also briefly touched [see Fig.~3(a) in this reference], 
but only for rather small system sizes, $L = 32$, such that the size-dependence
of the time needed to reach a
GS could not be well studied. As visible in Fig.~\ref{fig:p0_tmc_delta10}, 
the time needed to reach a GS with $p_0\approx 1$  depends
on the system size $L$, and longer times are needed for larger
sizes. Interestingly, the curves for
linearly growing system sizes are spaced more or less equidistantly on the
logarithmic time axis. This means that finding an exact GS becomes
exponentially harder as the lateral system size increases linearly.
Thus, the ``physical'' single-spin flip approach even when combined
with the SWAP dynamics 
is computationally slower than the exact
GS algorithm, which runs in polynomial time. 

In spite of the negative conclusion regarding running times, the resulting
SWAP ensemble is of interest on its own. In the next
subsection, we will study the effect the SWAP annealing has on the magnitude of
the frustration structure of the obtained ${\mathcal J}^*$ realizations.

\subsection{Level of Frustration}

\definecolor{blue_link}{rgb}{0.1, 0.2, 0.8}
\definecolor{red_link}{rgb}{0.8, 0.1, 0.1}
\definecolor{vertex_black}{rgb}{0.0, 0.0, 0.0}

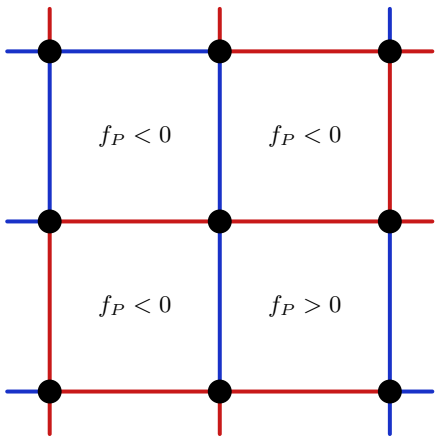
\begin{figure}[t!]

\definecolor{blue_link}{RGB}{25, 50, 200}
\definecolor{red_link}{RGB}{200, 25, 25}

\begin{center}

\begin{tikzpicture}[scale=2.25, line width=1.5pt, cap=round]


    \draw[red_link]  (0,2) -- (0, 2.25);   
    \draw[blue_link] (0,2) -- (-0.25, 2);  
    \draw[blue_link] (2,2) -- (2, 2.25);   
    \draw[red_link]  (2,2) -- (2.25, 2);   
    \draw[blue_link] (0,1) -- (-0.25, 1);  
    \draw[red_link]  (0,0) -- (0, -0.25);  
    \draw[red_link]  (2,1) -- (2.25, 1);   
    \draw[blue_link] (2,0) -- (2, -0.25);  

    \draw[red_link]  (1,2) -- (1, 2.25);   
    \draw[red_link]  (1,0) -- (1, -0.25);  
    \draw[blue_link] (0,0) -- (-0.25, 0);  
    \draw[blue_link] (2,0) -- (2.25, 0);   

    \draw[blue_link] (0,2) -- (1,2); 
    \draw[blue_link] (0,1) -- (0,2); 
    \draw[blue_link] (1,1) -- (1,2); 
    \draw[red_link]  (0,1) -- (1,1); 
    \node at (0.5,1.5) {$f_P < 0$};  

    \draw[red_link] (1,2) -- (2,2); 
    \draw[red_link] (2,1) -- (2,2); 
    \draw[red_link] (1,1) -- (2,1); 
    \node at (1.5,1.5) {$f_P < 0$};  

    \draw[red_link] (0,0) -- (0,1); 
    \draw[red_link] (0,0) -- (1,0); 
    \draw[blue_link] (1,0) -- (1,1); 
    \node at (0.5,0.5) {$f_P < 0$};  

    \draw[blue_link] (2,0) -- (2,1); 
    \draw[red_link]  (1,0) -- (2,0); 
    \node at (1.5,0.5) {$f_P > 0$};  

    \foreach \x in {0,1,2} {
        \foreach \y in {0,1,2} {
            \fill[black] (\x,\y) circle (2pt);
        }
    }

\end{tikzpicture}

\end{center}
\caption{Four plaquettes in the square lattice. Blue and red links represent positive and negative 
couplings, respectively. The sign of the local frustration $f_P$ is indicated within each plaquette. Neighboring 
plaquettes share a vertical or horizontal bond. The two upper plaquettes and the lower-left one 
belong to a cluster of frustrated plaquettes with $f_P<0$, while the lower-right one could be part 
of a cluster of highly unfrustrated plaquettes with neighboring ones to the right and below
if their $f_P>f_{\rm thres}$.}
\label{fig:sketch}
\end{figure}

The main purpose of the SWAP algorithm is to reduce the frustration in the
system by, in general,
reducing the magnitude of the bonds ${\mathcal J}_{ij}$ contained in
frustrated plaquettes, while bonds ${\mathcal J}_{ij}$ in
non-frustrated plaquettes acquire,  on average, larger magnitudes.
This is implemented by SWAP moves between pairs of spin lengths
which are associated to the bonds through the Hamiltonian.
The SWAP moves preserve the average spin length but not
the resulting average bond weight~\cite{miranda2024}.
The process target of reducing the energy is  not trivially reachable,
because each bond belongs to two plaquettes and each spin is reached by four bonds,
and many bonds appear in frustrated and
non-frustrated plaquettes as well, see the sketch in Fig.~\ref{fig:sketch}.

\begin{figure}[t]
\begin{center}
    \includegraphics[width=0.9\columnwidth]{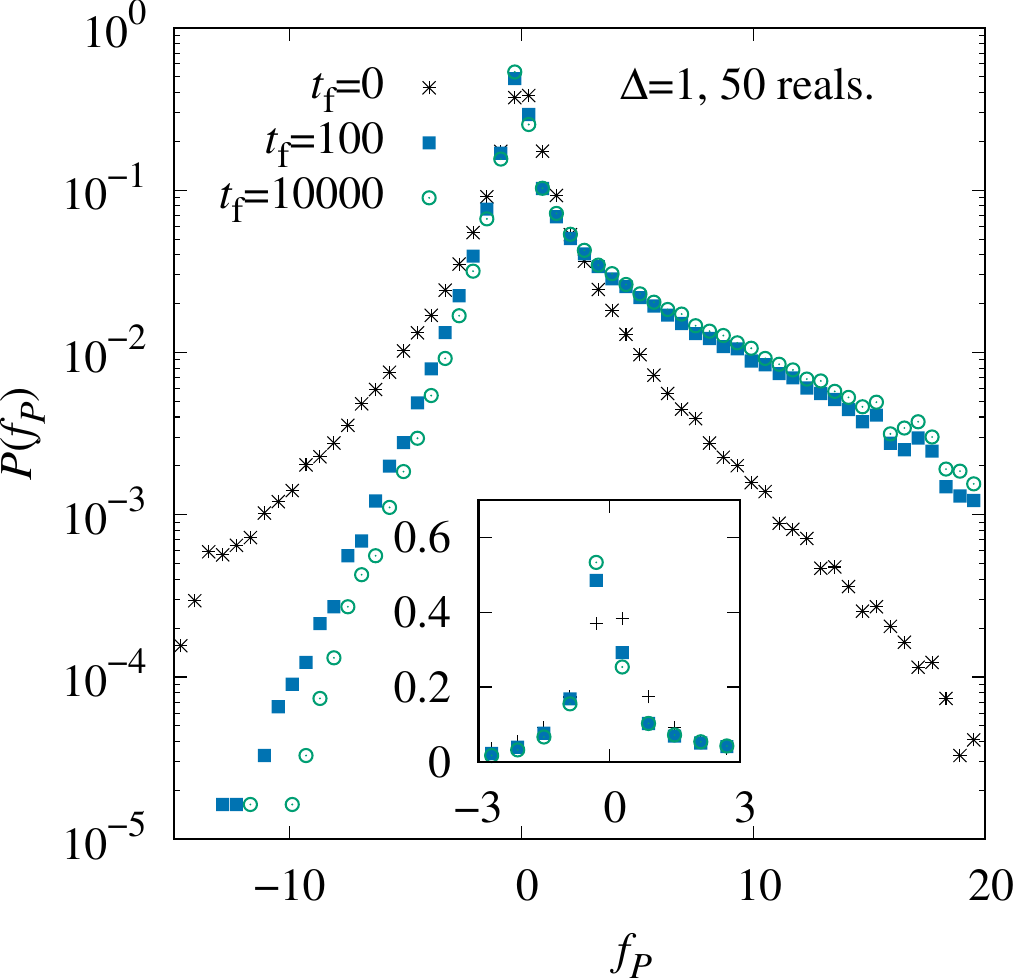}
\end{center}
\vspace{-0.25cm}
\caption{\label{fig:frust_distr_delta10}
      Distribution of plaquette frustration values $f_{P}$ defined in Eq.~(\ref{eq:f:P})
      for $\Delta=1$ and 
      $L=64$, using the square annealing procedure in Eq.~(\ref{eq:annealing}) with 
      $t_f=0$, $t_f=100$ and $t_f=10^4$ SWAP
      annealing sweeps, respectively. The distributions were sampled using 50 realizations of the initial disorder. Inset: zoom over the central part of the distribution
      in double linear scale.
      }
\end{figure}

  To investigate this issue, we study the resulting
  distribution of plaquette frustration values $f_P$ defined
  in Eq.~(\ref{eq:f:P})~\footnote{In the SM of~\cite{miranda2024} the local frustration 
  was analyzed. Figure SM25 shows
the time evolution of the number density of plaquettes with strong $|f_P|>1$
and weak  $|f_P|<1$ negative frustration, and 
it was concluded that ``the number density of plaquettes with small frustration dominates over the ones with large frustration asymptotically''. From the analysis of Fig. SM29 it was observed that 
``In the course of time the symbols rearrange spatially. Rather large regions with small frustration are visible in the late snapshots.''}. The probability distribution functions for 
  50 realizations of a system with $L=64$, using 
  $\Delta=1$ and different annealing times $t_f$ 
  are shown in Fig.~\ref{fig:frust_distr_delta10}. Initially, without annealing, i.e.,
$t_f=0$, the distribution is symmetric, most plaquettes exhibit
a frustration $f_P$ near 0. After $t_f=100$ annealing
steps, very negative frustration values have disappeared; instead, very positive
ones, like $f_P> 20$, become much more likely. This matches the fact that
the SWAP algorithm 
tries to decrease the energy and therefore
the frustration. This feature is slightly more visible for a much slower
annealing, datapoints for $t_f=10^4$, although the form of the 
distribution has not changed much.
  Thus, finding a GS becomes more likely with slower annealing,
  as we have seen in the previous section, but the distribution of the
  local frustration values of the
  bond-disorder realizations are
  not changed much as compared to the faster
  SWAP annealing.

    \begin{figure}[t]
\begin{center}
    \includegraphics[width=0.35\textwidth]{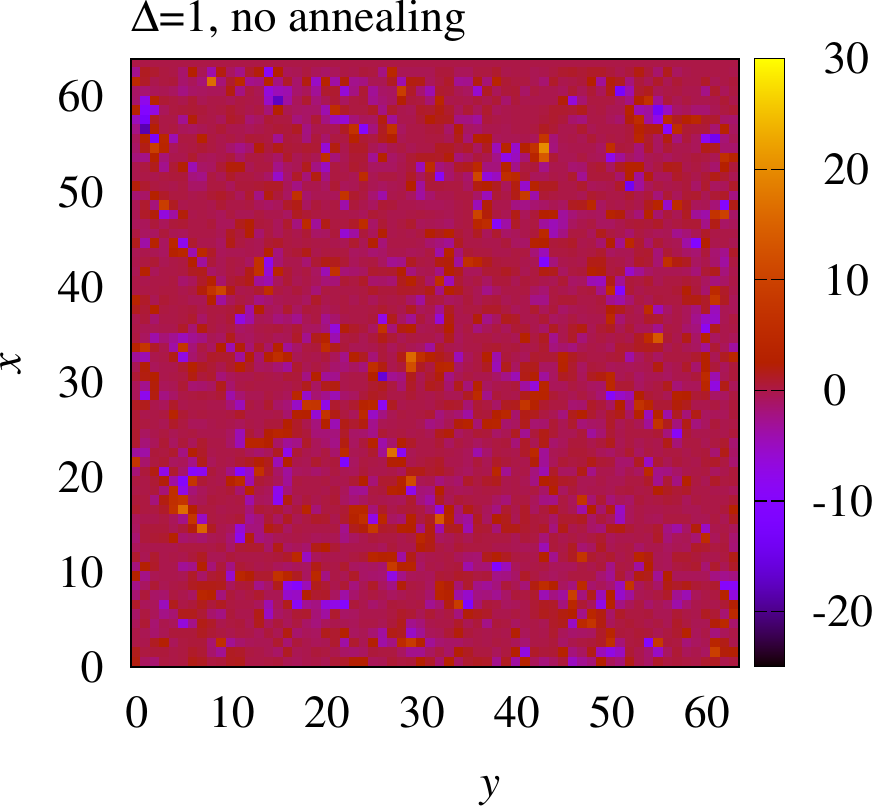}
\includegraphics[width=0.35\textwidth]{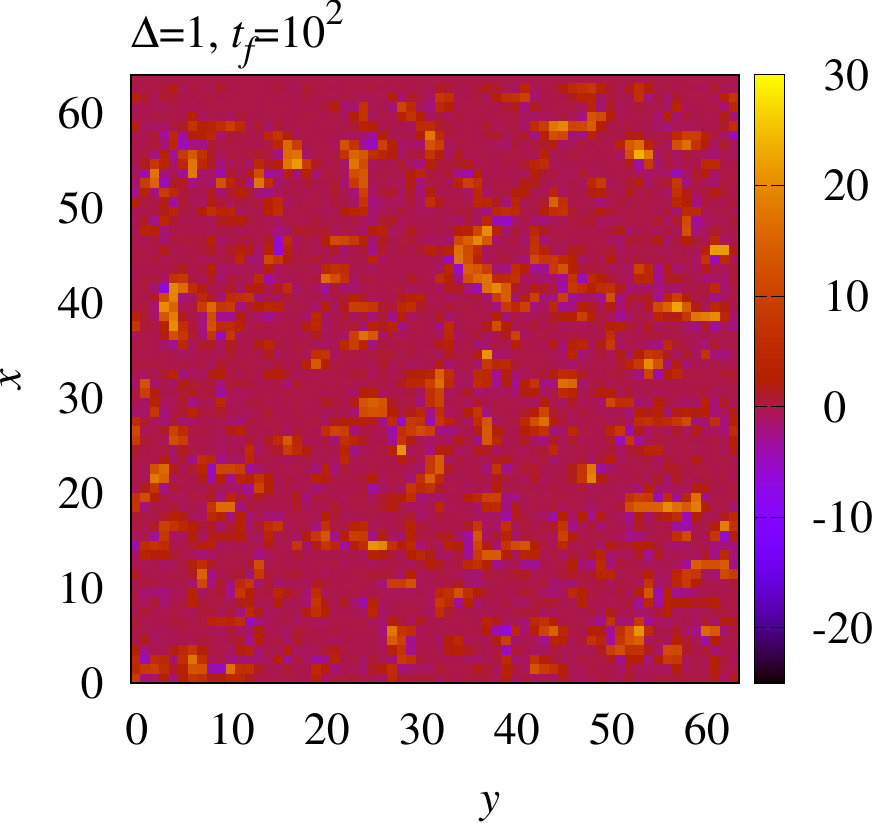}
\includegraphics[width=0.35\textwidth]{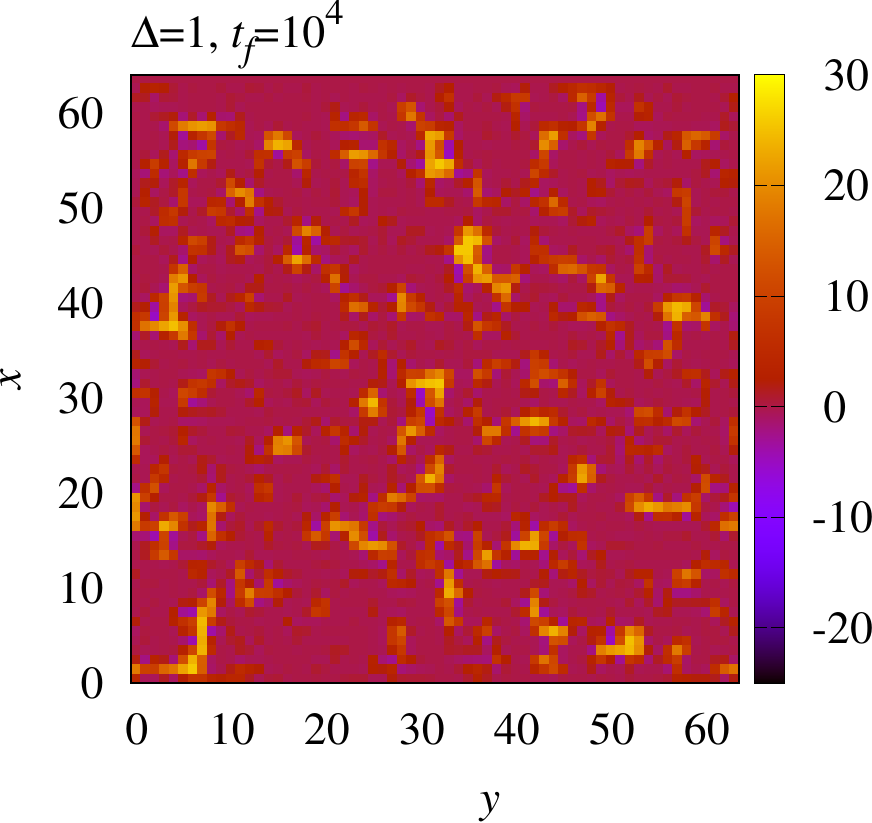}
\end{center}
\vspace{-0.25cm}
\caption{\label{fig:frustration_pattern}
      The set of plaquettes frustration value $f_{P}$ in a system with 
      $L=64$ lateral size,  $\Delta=1$, and 
      three cases: no annealing (top), quadratic annealing with $t_f=10^2$ annealing steps (middle)
      and $t_f=10^4$ annealing steps (bottom).
    }
\end{figure}

Examples for the actual spatial distribution of $f_P$  can be seen
in Fig.~\ref{fig:frustration_pattern}. While for $t_f=0$ the structure
seems to be completely random, which it is, one sees the emergence of
small clusters with enhanced $f_P>0$  (yellow color)
for $t_f=100$. This trend continues, but does not influence much the 
snapshots, for $t_f=10^4$. These clusters accumulate bonds with larger
magnitudes at those spatial locations,
where many unfrustrated plaquettes are present by random
fluctuations. Note that each bond is present in exactly two plaquettes.
Therefore, the local accumulation of large bonds in clusters of unfrustrated
plaquettes often results in few adjacent frustrated clusters  with
also a large magnitude of $|f_{P}|$. This is well visible in the plot by few
violet plaquettes integrated into the highly unfrustrated clusters.

\begin{figure}[htb]
\vspace{0.5cm}
\begin{center}
    \includegraphics[width=0.9\columnwidth]{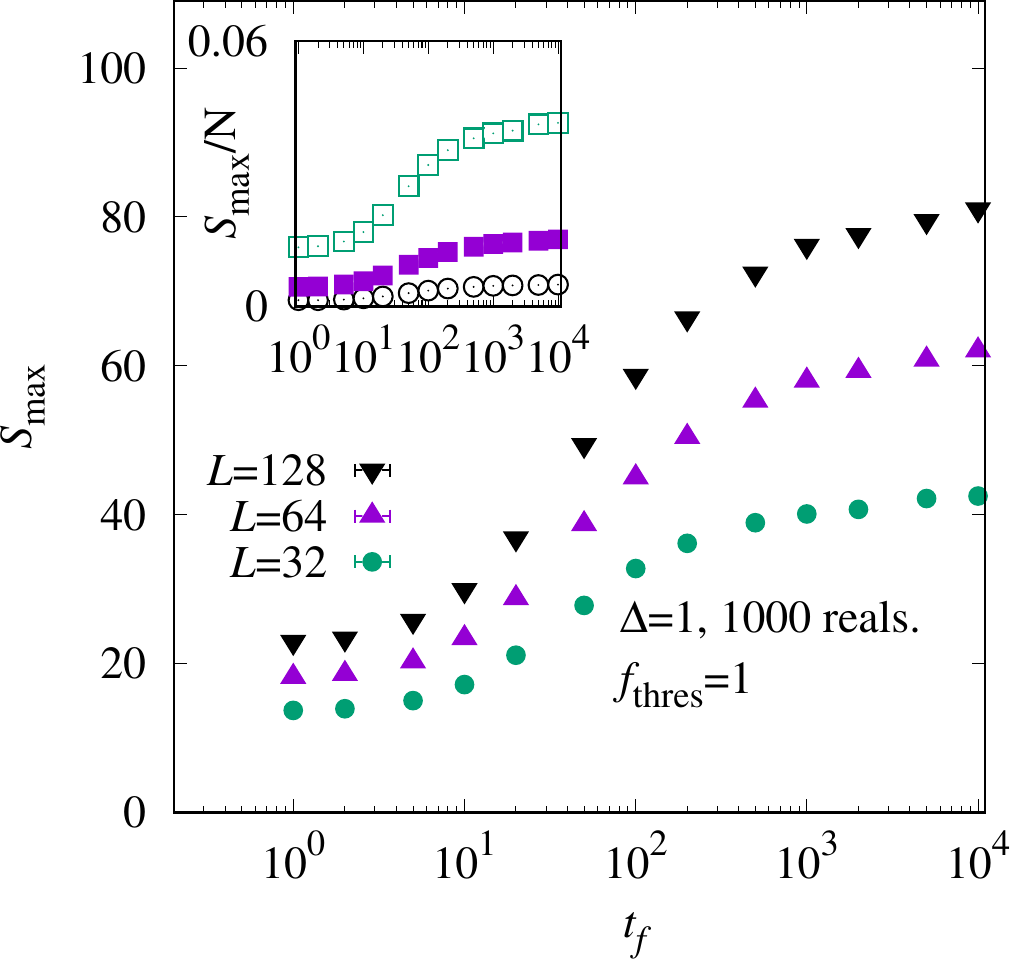}
\end{center}
\vspace{-0.25cm}
\caption{\label{fig:cluster_tmc_delta10}
  Average size $S_{\max}$ of the largest cluster of highly unfrustrated plaquettes
  as a function of the annealing time $t_f$ for $\Delta=1$ and three
  system sizes $L=32,64$ and $128$. The inset shows the relative size
  $S_{\max}/N$ as a function of $t_f$.
}
\end{figure}

To analyze these clusters, we define a \emph{highly unfrustrated} (HU)
plaquette as
a plaquette $P$ where $f_{P}$ is larger than some threshold value
$f_{\rm thres}$. We use, somehow aribtrary, $f_{\rm thres}=1$
for the case $\Delta=1$, which is clearly larger than
the typical value of $f_{P}$ for unfrustrated plaquettes  as visible
in Fig.~\ref{fig:frust_distr_delta10}. The results do not change
qualitatively for other choices of $f_{\rm thres}$ provided that it is
 significantly larger than zero. We define two plaquettes
 as \emph{neighbours} when they share a bond, see Fig.~\ref{fig:sketch}. 
 Then, clusters are maximum-size
 sets of neighbouring HU plaquettes, i.e.,
the transitive closure of the neighbour relation for HU plaquettes,
as in standard percolation. Note that, since the plaquettes are
sites on the dual lattice and the frustration values $f_{P}$ are
assigned to its sites, this definition corresponds to the site percolation model,
not the bond percolation one. We determine the clusters, called connected
components in graph theory, using a standard
depth-first search \cite{cormen2001} and measure the size  of the largest
HU cluster for each realization, resulting in an average maximum size
$S_{\max}$.

In Fig.~\ref{fig:cluster_tmc_delta10} we show $S_{\rm max}$
as a function of the number $t_f$ of annealing sweeps. In the beginning,
  by chance, there are already some clusters present, as also visible
  in Fig.~\ref{fig:frustration_pattern}. With growing annealing time,
  the maximum cluster size increases, but not very much. In particular, the
  cluster size does not become extensive: by doubling the lateral system
  size from $L=32$ first to $L=64$ and again to $L=128$,
  corresponding to multiplying the number of spins twice by a factor
  of four,  $S_{\max}$ does not increase by a factor of 16, as it would be
  the case for an extensive quantity. Actually,
  for $L=128$, which corresponds to 16384 spins, the largest cluster
  occupies only a fraction of about 0.006, see also
  the inset of Fig.~\ref{fig:cluster_tmc_delta10}. Thus, the swapping of the
  spin lengths and  the resulting changes of the bond magnitudes 
  do not lead to the formation of large clusters of HU
  plaquettes.  Clearly, a full percolation transition cannot occur because the frustrated plaquettes 
  remain frustrated under the SWAP moves; that is, the sign of $f_{P}$ does not change.
   With 50\% of all plaquettes being
  frustrated and the site percolation threshold in two dimensions
  being  about 0.59~\cite{stauffer1994}, the set of unfrustrated plaquettes will never percolate
  with notable probability.

\begin{figure}[htb]
\vspace{0.5cm}
\begin{center}
    \includegraphics[width=0.9\columnwidth]{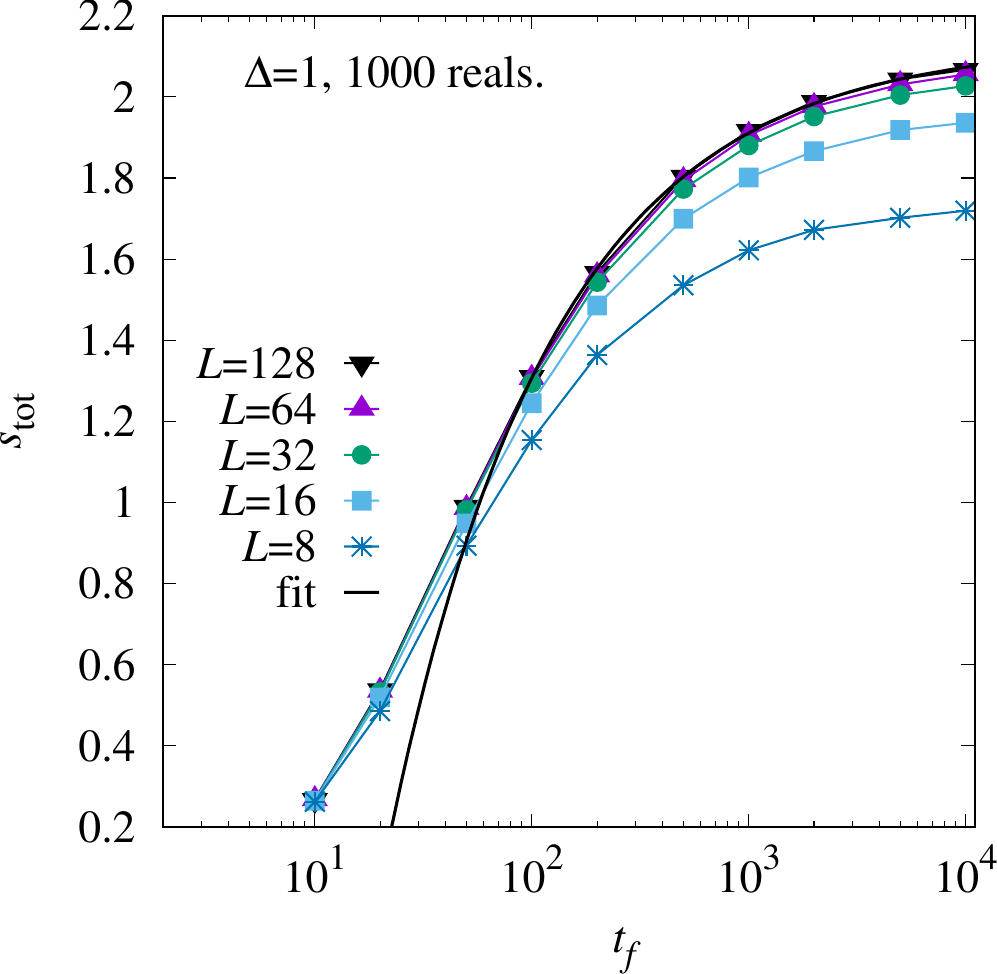}
\end{center}
\vspace{-0.25cm}
\caption{\label{fig:ftot_tmc_delta10}
      The average total satisfaction $s_{\rm tot}$ as a function of
      the number $t_f$ of annealing steps, for
      $\Delta=1$ and  different system sizes $L$. Lines are guides to the
      eyes only
      except for the black curve which  represents
      a fit of the $L=128$ data to the power-law in Eq.~(\ref{eq:stot:fit}).
}
\end{figure}

Since, as we have just seen, the properties of a realization will
not be determined by large clusters of unfrustrated plaquettes,
we consider an overall view on the level of frustration in the annealed
realizations. We show
in Fig.~\ref{fig:ftot_tmc_delta10} the average total satisfaction $s_{\rm tot}$
defined in Eq.~(\ref{eq:stot:def}) as a function of the annealing steps
$t_f$. For very small system sizes, some finite size effects
at longer annealing times are visible, but for $L\ge 32$ the result is
almost independent of the system size. This is in contrast to
the above studied probability $P_{0}$ to obtain a GS, see
Fig.~\ref{fig:p0_tmc_delta10}.

For longer annealing times $t_f \ge 100$
one can fit rather well a power-law function which models the
approach to the infinite-time annealing limit via
\begin{equation}
  s_{\rm tot}(t_f) = s_{\rm tot}^{\infty} + b\,t_f^{-c}\,.
  \label{eq:stot:fit}
\end{equation}
where for convenience, we denote by $b$
 and $c>0$  the correction fitting parameters. For $b<0$ the approach to the limiting value $s_{\rm tot}^{\infty}$ takes place 
 from below.
For $L=128$ we have obtained $s_{\rm tot}^{\infty}=2.13(1)$ where at
$t_f=10^4$ already a value of $s_{\rm tot}=2.07$ was attained for
this system
size. This means that after $10^4$ steps, the total satisfaction does not change much
for all system sizes,
in contrast to the GS probability $P_{0}$,
which changes still considerably
for a rather large system $L=128$, see Fig.~\ref{fig:p0_tmc_delta10}. 
The vertical shift with system size which is visible in Fig.~\ref{fig:p0_tmc_delta10} indicates that this will hold also for even larger system sizes.

\subsection{Domain-Wall Energy \label{sec:domain_wall}}

In order to see whether a system exhibits order at finite temperature,
and what type of order it is, we consider
the dispersion $\sigma_{\Delta E_0}$
of the domain wall energies $\Delta E_0$, 
as defined in Eqs.~(\ref{eq:de0}) and (\ref{eq:def:width}).

\subsubsection{Size dependence for long annealing time}

\begin{figure}[b!]
\vspace{0.5cm}
\begin{center}
\includegraphics[width=0.9\columnwidth]{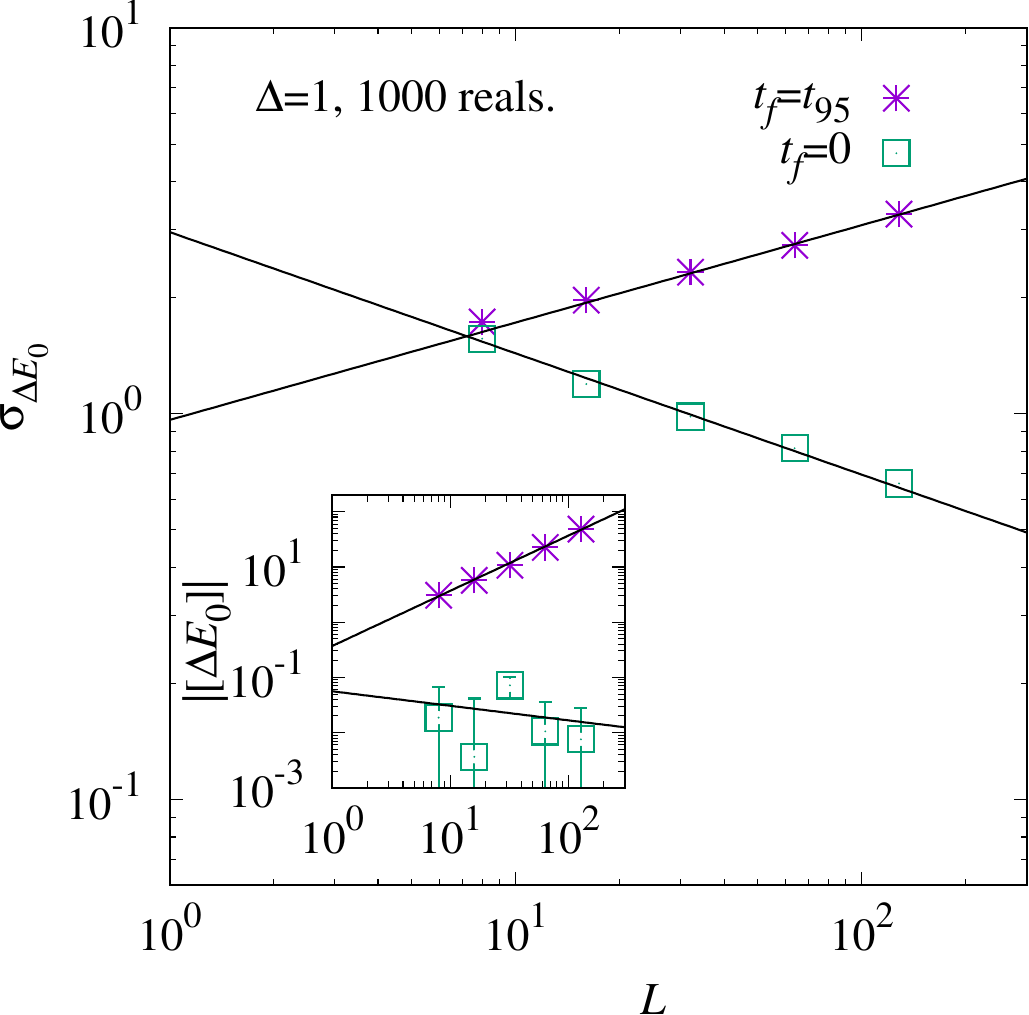}
\end{center}
\vspace{-0.25cm}
\caption{\label{fig:stiff_L_Delta1}
The width $\sigma_{\Delta E_0}$
of the domain-wall energy distribution as a function of the system size $L$,
for $\Delta=1$ and
annealing times $t_f=t_{95}=t_{95}(L)$ such that for 95\% of all
realizations GSs are found.
For comparison, the case without annealing, i.e. $\Delta=1$ and $t_f=0$
is also shown. The inset displays the absolute mean value
$|[ \Delta E_0  ]|$ 
also as a function of the system size $L$. Lines
represent fits to the power laws in Eqs.~(\ref{eq:def:theta}) and
(\ref{eq:mean:E0:scale}), respectively.
}
\end{figure}

Since the motivation of the SWAP approach was to obtain ground states,
we start with the case in which the number $t_f$ of annealing
sweeps is chosen 
such that for  almost  all realizations, i.e., for 95 \%,
a GS is found by the SWAP annealing.
We call this time $t_{95}$, which can be read off from
Fig.~\ref{fig:p0_tmc_delta10} for $\Delta=1$.
We restrict ourselves
to $L\le 128$ since otherwise the number of annealing steps becomes
extremely large. The obtained realizations of
annealed bonds ${\mathcal J}^*$ are stored. For these realizations
the matching approach is used to obtain exact ground states for periodic
as well as antiperiodic boundary conditions, as described in
Sec.~\ref{sec:domain-walls}. This yields a sample of
domain-wall energies $\Delta E_0$, see Eq.~(\ref{eq:de0}).
 The resulting width
 $\sigma_{\Delta E_0}$ of the distribution of $\Delta E_0$, as
 defined in Eq.~(\ref{eq:def:width}),
 is shown as a function of the system size $L$ in Fig.~\ref{fig:stiff_L_Delta1}.
 For comparison, the corresponding
 result for the case in which the realizations have been obtained
 with random lengths but without annealing,
 i.e.\ $t_f=0$, is included as well.
 
One observes a clear growth
of the width for $t_f=t_{95}$, which indicates strong fluctuations of the domain-wall
energy. More importantly, as
shown in the inset, the absolute mean value 
increases like a power law
\begin{equation}
  \label{eq:mean:E0:scale}
  |[ \Delta E_0  ]| \sim L^{\tilde \theta}
\end{equation}
with exponent $\tilde \theta=1$, i.e., linearly
with the system size, which is exactly the behavior of a ferromagnet.

\begin{figure}[htb]
\begin{center}
\includegraphics[width=0.7\columnwidth]{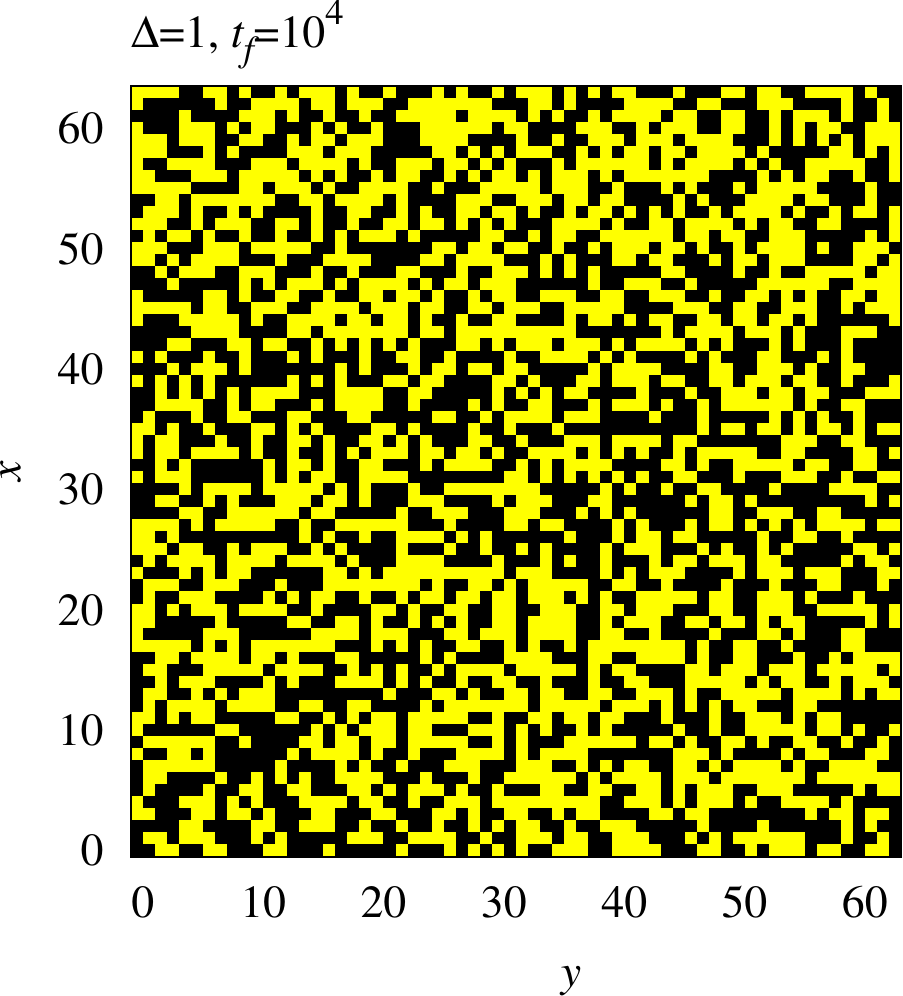}
\end{center}
\vspace{-0.25cm}
\caption{\label{fig:config_tmc10000}
An example spin configuration obtained for a system of size $L=64$ and $\Delta=1$ 
obtained after $t_f=10^4$ annealing sweeps.}
\end{figure}

On the contrary, for the case $t_f=0$, where the bonds exhibit
scaled values but
no swaps have been performed, both the width $\sigma_{\Delta E_0}$
and the absolute mean
value  decrease as functions of the
system size $L$.
This is exactly the behavior of a standard two-dimensional spin-glass.
More precisely, for $|[ \Delta E_0  ]|$ we have found
$\tilde \theta=-0.31$. This value of the exponent 
describes  the reduction of the statistical fluctuations of the domain-wall energy, center about zero, 
due to self-averaging when increasing the system size.
Note the large error bars of the mean values in the
inset of the figure, which often include
zero.
The width follows 
a power law with $\theta=-0.25(1)$ for $L\ge 32$, which is,
considering the small sizes used here,
comparable to the value of the
scaling of the
domain-wall energy as found in
two-dimensional spin-glasses \cite{stiff2d,aspect-ratio2002,khoshbakht2018}.
Thus, merely rescaling the bonds
does not destroy the $T=0$ spin-glass behavior. But after
performing the SWAP annealing long enough such that almost always
GSs are found, the system has lost its spin-glass property.
Note that the realizations  do not exhibit a notable
magnetization, because the SWAP annealing does not change the
signs of the bonds and the positive bonds cannot grow
strong enough to favor a nonzero global magnetization. We have 
also confirmed the absence of a significant magnetization
in our simulations.
As an example, we show a spin configuration $\sigma^*$ for $L=64$, $\Delta=1$ obtained after $t_f=10^4$ annealing sweeps in Fig. \ref{fig:config_tmc10000}. The configuration is basically random and contains no significant 
spin-spin correlation. We have explicitly verified this fact by calculating  
the spin-spin correlation $\langle \sigma^*_i \sigma^*_j \rangle_r - (m^*) ^2$, for the configurations $\sigma^*$ obtained after annealing during $t_f=10^4$ steps and averaged over 1000 realizations of systems with $L=64$. The average $\langle \ldots \rangle_r$ is over all spins separated by a spatial distance $r$  and $m^*$ is the magnetization $\sum_i\sigma_i/N$ of $\sigma^*$. As a result, for any nonzero separation $r$ of the spins $\sigma^*_i$ and $\sigma^*_j$ the measured correlation was statistically compatible with zero. 
The reason is that the ferromagnetic order is hidden by the implicit gauge transformation. One could use the obtained ground states to find the gauge transformation which maximizes the magnetization. But this procedure would only highlight the large fraction of satisfied bonds in the ground state and would also yield the same result for true spin glasses. Thus, the approach of calculating the domain-wall energies used here is better suited to identify hidden ferromagnetic order and distinguish it from spin-glass order.

To summarize the domain-wall energy results, the SWAP algorithm
has turned the spin-glass realizations
into gauge-transformed ferromagnets. Note that although the magnetization is zero, the system is not close to criticality, 
because otherwise the mean of the domain-wall energy, for long-enough annealing time, would not grow with size.
Thus, the system has become in some sense
physically ``easy'', although finding the GS with 
SWAP dynamics still takes a time growing exponentially with system size.

\begin{figure}[htb]
\vspace{0.5cm}
\includegraphics[width=0.9\columnwidth]{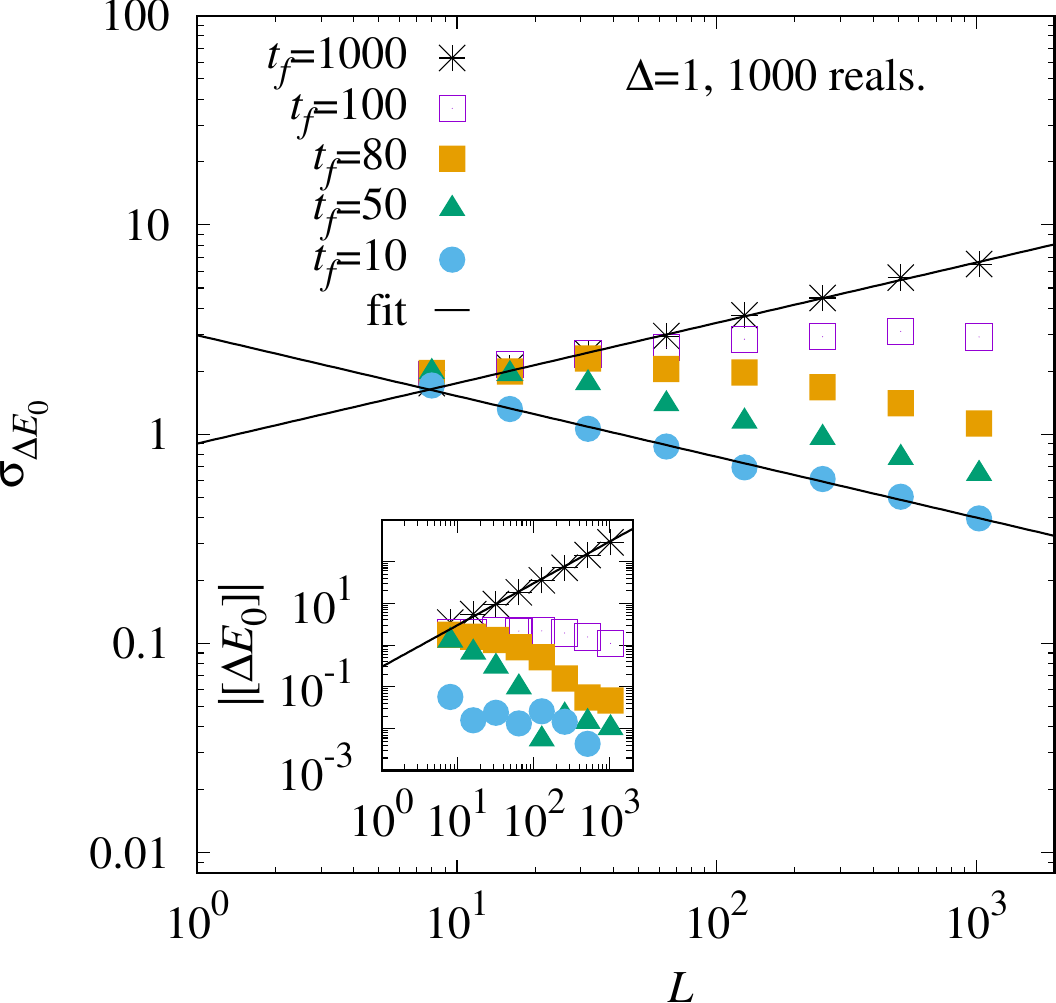}
\caption{\label{fig:de0_delta10_L_tmc}
The width $\sigma_{\Delta E_0}$
of the domain-wall energy distribution as a function of the system size $L$,
for different annealing times $t_f$. The inset shows the absolute
mean  value      $|[ \Delta E_0 ]|$ of
the distribution as a function of the system size $L$.
}
\end{figure}

\subsubsection{Effect of the annealing time $t_f$}

The next question that we consider is whether these
results actually depend on
the fact that the SWAP algorithm was run long enough to almost
always find a GS of the Hamiltonian Eq.~(\ref{eq:H:annealed}).
In the previous section we have seen that, for a given value of $\Delta$
and large enough system size, the average frustration
only depends on the annealing time $t_f$ and not on the system size.
Since the frustration
of a spin-glass is the main quantity that determines its physical behavior,
here we select the bond realizations ${\cal J}^*$ 
directly according to the annealing time $t_f$. 
For each value $t_f$ we consider
a large set of system sizes. This also means that we ignore the small shift of
$s_{\rm tot}$ which is visible for small system sizes.
We repeat this domain-wall analysis  for realizations ${\cal J}^*$ obtained for several values of $t_f$, respectively.
This procedure allows us to reach
larger system sizes, because we do 
not require a high probability of reaching a ground state, thus we do not need extremely long annealing times in this way.

In Fig.~\ref{fig:de0_delta10_L_tmc} the
width $\sigma_{\Delta E_0}$
of the domain-wall energy distribution as a function of the system size
$L$ up to size $L=1024$ is shown
for different annealing times. In the inset the
average of this distribution is displayed. One observes that
for a small value of
$t_f=10$ the behavior is similar to the one of the non-annealing case:
the average domain-wall energy $[\Delta E_0]$ is about zero,
more and more with increasing system size,
but fluctuates strongly such that a fit does not make sense.
The width of the
distribution decreases like Eq.~(\ref{eq:def:theta}). When fitting for $L\ge 32$ 
we have found $\theta=-0.288(11)$ which is compatible with the known
results~\cite{stiff2d,aspect-ratio2002,khoshbakht2018}
for two-dimensional spin-glasses. The width results
for $t_f=50$ and $t_f=80$ exhibit the same slope 
in the log-log plot for large-enough system sizes. The average is also small
and approaches zero more and more. Thus, for $t_f \le 80$
the typical behavior of the two-dimensional spin-glass
remains.

For a SWAP annealing time about $t_f=100$ the behavior changes:
Roughly, both mean and width do not depend on the system
size any more. When studying longer annealing times, here we present
$t_f=1000$, we recover the behavior of the ferromagnet:
The mean of the domain-wall energy
grows linearly with system size, i.e., $\tilde \theta=1$,
while the fluctuations have a 
growth, with a power-law with exponent $\theta=0.290(8)$. 

Thus, a relatively small number $t_f$ of SWAP annealing sweeps is sufficient to map the spin-length-scaled two-dimensional spin glass onto a gauge-transformed ferromagnet. This holds true even though, for such a small number of sweeps, the resulting spin configurations $\sigma^*$ are typically not ground states of the bond realizations $J^*$ (i.e., $P_0 \approx 0$), except for very small system sizes $L$.

\begin{figure}[htb]
\vspace{0.5cm}
\includegraphics[width=0.9\columnwidth]{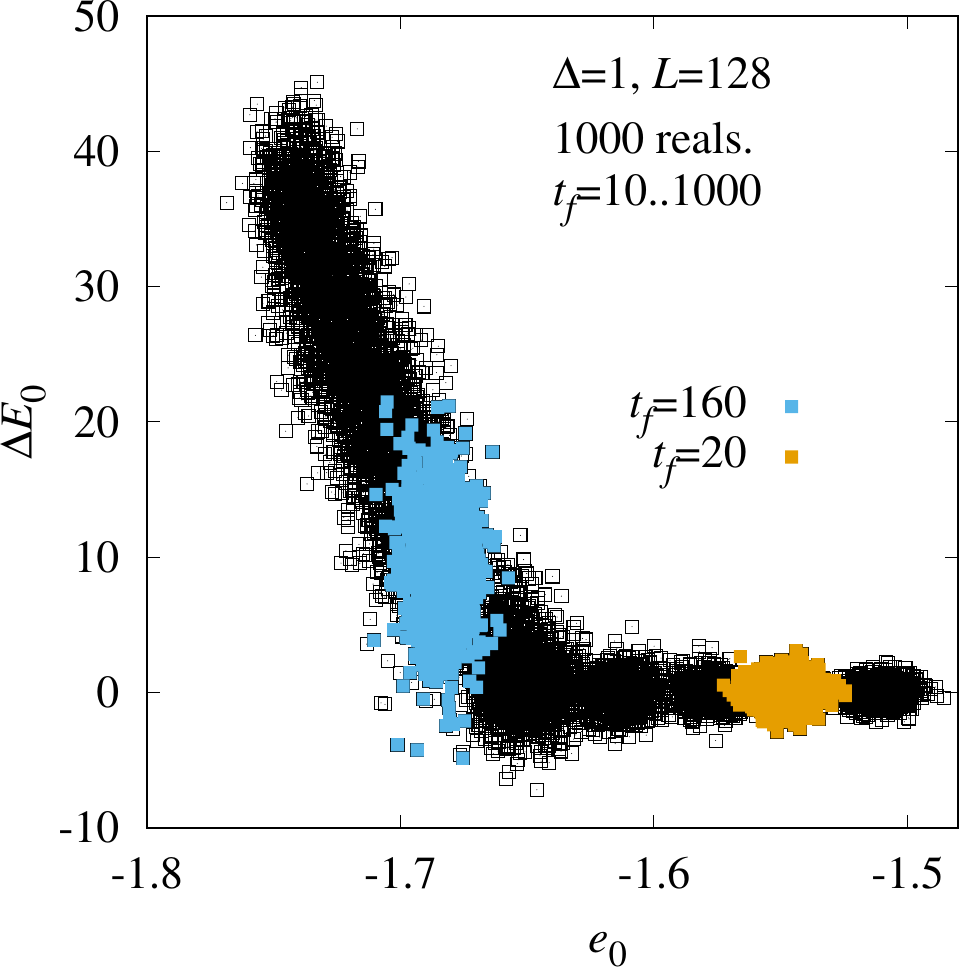}
\caption{\label{fig:corr_e0_de0_delta10}
  A realization-wise scatter plot of the domain-wall
  energy $\Delta E_0$ as a function of the GS energy density
  $e_0$ for $\Delta=1$ and $L=128$. The data for 11 different
  annealing times $t_f=10,20,$ $30,50,$ $80,100,$ $160,200,$ $300,500,$ and $1000$ 
  are lumped together.
  The data points contributed by $t_f=20$ are highlighted in orange and those by $t_f=160$ in light blue.
}
\end{figure}

\subsubsection{Correlation between the ground state energy and its dispersion}

The GS energy density $e_0$
of a system is related to its physical behavior.
A low GS energy typically means a ferromagnetic behavior while
spin-glasses exhibit higher values of $e_0$. 

We have shown that the magnitude of the couplings as defined in Eq.~(\ref{eq:var:J:delta}) does not change much by rescaling with the spin lengths, over the full range $\Delta \in [0,2]$. But the exchange of spin length values and corresponding changes of the couplings lead to much lower GS energies, see Fig.~\ref{fig:e0_L}. Thus, changes in the GS energy may be related to changes of the physical behavior.
To investigate whether this
is visible for the annealed SWAP ensemble as well, we present in
Fig.~\ref{fig:corr_e0_de0_delta10} a realization-wise scatter plot
of the results for the GS energy density $e_0$ and the observed
domain-wall energy $\Delta E_0$. The results were obtained for 1000
realizations of size $L=128$ with $\Delta=1$ and 11 different
SWAP annealing times $t_f \in [10,1000]$.
One observes a clear correlation
between $e_0$ and the domain-wall energy distribution for small values
of $e_0$ below $-1.65$. Here, the average domain-wall energy grows and
the scatter
of the data points  increases when $e_0$ becomes smaller.
This is compatible with the results
for the dependence of the average and width of the distribution of
$\Delta E_0$ shown in Fig.~\ref{fig:de0_delta10_L_tmc}.

On the other hand, for larger
values of the GS energy density, the domain-wall energies
are distributed around 0, independently of $e_0$.  Thus, no
correlation is visible here. Only the spread of the data points
slightly decreases with increasing GS energy.

The observed correlation for small energies
is mainly due to the fact that
results obtained from different values of $t_f$ are collected
within one plot. This can be seen by considering the results obtained
for a single value of the SWAP annealing time, e.g., $t_f=160$,
  which is highlighted in light blue in the figure. For these datapoints, 
  no strong  structure is visible,
  and the measured correlation coefficient $-0.11$ is near zero. This becomes
  also visible in the figure
  for smaller values of $t_f$ which appear at higher
  values of $e_0$ in the right of the plot, where one can see that the data
  is a superposition of few rather round clouds of data points  
  (see the datapoints highlighted in orange which represent the result for $t_f=20$).
  Thus, the change
  of the behavior is driven by the SWAP annealing time spent on each
  realization rather than the actual fluctuations of the GS
  energy which is, as we will also see in the next section,
  related to the fluctuations of the total satisfaction.

\subsection{Physical Hardness \label{sec:hardness}}

Since for the planar Ising spin-glass the GS can be computed
in polynomial time, it is an ``easy'' computational problem. Nevertheless,
when using local single-spin flip annealing, which corresponds to the 
physical dynamics, this changes. As we have seen in
Sec.~\ref{sec:gs_probability}, it takes a typical time which increases
exponentially with the system size to find a GS, even
when involving the non-local SWAP moves. To measure a physical hardness,
we now consider the case of
an annealing with just single-spin flips, without the SWAP moves, as
a  second set of annealing runs. Thus, pure
physical local dynamics or, in other words, updates with $p_{\rm swap}=0$. 
We apply this second-type
annealing  to different kinds of realizations,
the original ones ($\Delta=0$) and those obtained and
stored when including
different lengths ($\Delta>0$) without ($t_f=0$)
and with ($t_f>0$) SWAP annealing. 
Note that for $L=16$ and $t_f=10^3$ in about 95\% of all cases a GS is found by the SWAP annealing, see Fig.~\ref{fig:p0_tmc_delta10}. But we do not distinguish whether in the SWAP annealing a GS was found or not, all realizations are considered.
For any of these three types
of stored realizations,
we perform 100 independent
second-type annealing runs.

We start from $T_0=1$ and decrease the temperature
down to $T=0$ with the quadratic
schedule of Eq.~(\ref{eq:annealing}).
To select the number of annealing sweeps
$t_{2,f}$, we performed
some tests. We observed
that finding GSs with only the local annealing
is much harder than with the SWAP annealing.
Thus,  we consider only a small system size, $L=16$,
and apply a rather  large number of 
annealing sweeps of the second type, $t_{2,f}=2\times10^5$, such that at least
some  GSs are found.
For even larger system sizes,
we would need many more sweeps $t_{2,f}$, without changing
the qualitative results that we present below.

\begin{figure}[ht]
\vspace{0.5cm}
    \includegraphics[width=0.9\columnwidth]{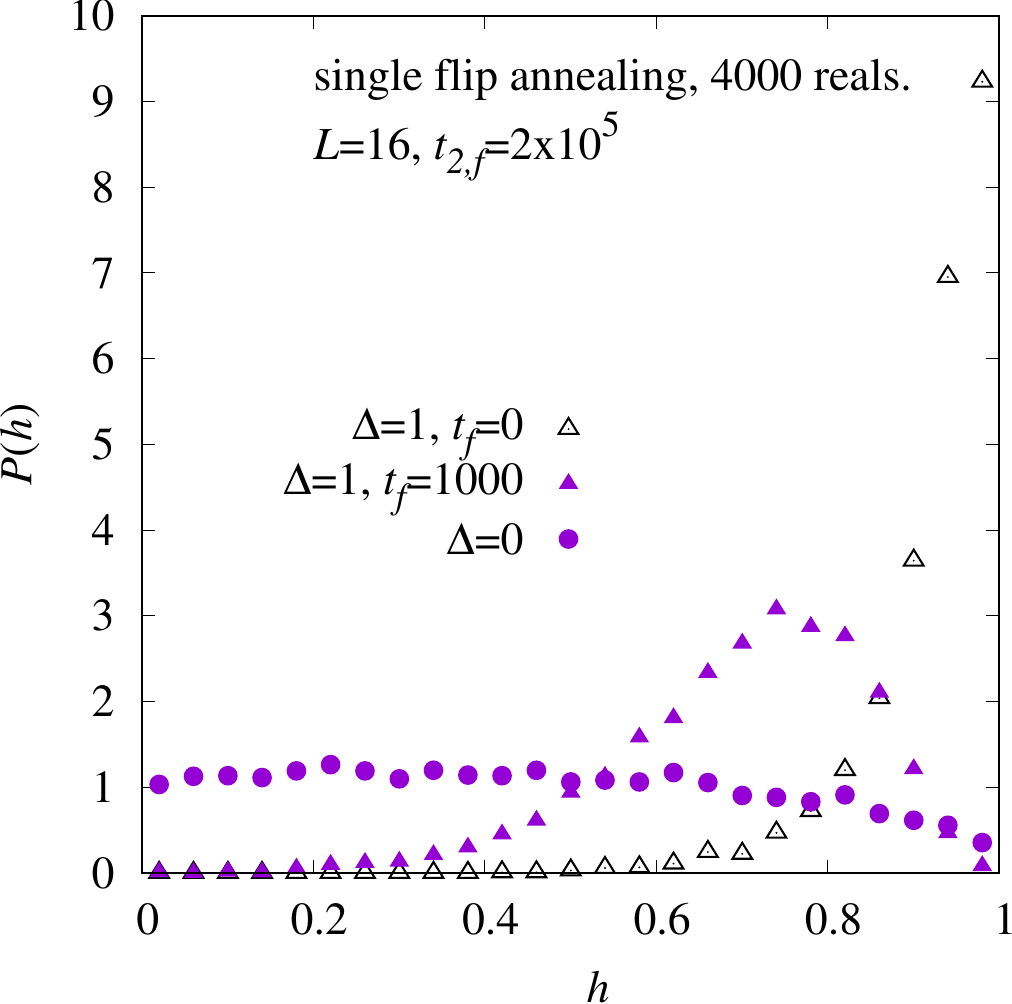}
    \caption{\label{fig:hardness_L16}
      The distribution of the hardness values $h$ for $L=16$ and
      the case of realizations with scaled bonds
      which were previously obtained
      without SWAP annealing ($\Delta=1$, $t_f=0$),
      with SWAP annealing ($t_f=1000$), and the original bonds
      ($\Delta=0$). 
      }
\end{figure}

We denote by  $p_{2,0}$, the
frequency how often a GS is found by the second
type of annealing, where the GSs
were again obtained by the matching approach for comparison.
If $p_{2,0}$ is small, GSs are hard to find, we therefore 
call $h=1-p_{2,0}$ the hardness.
The empirical distribution of the hardness, here obtained
from 4000 realizations for each type, is shown in
Fig.~\ref{fig:hardness_L16}. First, we discuss  the two cases with $\Delta=1$.
Without previous annealing, i.e., $t_f=0$, the hardness values are
rather close to one, i.e., GSs are hard to find. 
For the annealed
realizations with $t_f=1000$, the distribution of $h$ moves
to the left.
This means, that these realisations became
easier to treat also with the physical single spin-filp dynamics. This is reasonable, because the realizations are essentially
ferromagnets, as we have seen in Sec.~\ref{sec:domain_wall}. On the other
hand, for a pure ferromagnet
it is even easier to obtain the GS
by single-spin flip  annealing. For a $L=16$ ferromagnet
we found that already for a short annealing time of
$t_{2,f}=10^3$ in $p_{2,0}=62\%$ of all cases
a GS was found, which increased to $p_{2,0}=95\%$
for $t_{2,f}=10^4$, which is much smaller than the number
of sweeps needed for the SWAP realizations.
Thus, because the bond values are randomized and half of the plaquettes remain frustrated, the system stays quite challenging for single-spin-flip annealing, even when exhibiting ferromagnetic behavior.
Note that within the SWAP annealing, $t_f=1000$
leads in about $95\%$ of all cases to a ground state, thus when
creating these instances which involves changing the spin lengths, hence the bonds, most of the time a GS was found.
However, when we subsequently applied the second type of annealing—using only local steps, no length changes, and a much larger number of sweeps ($t_{2,f}=2\times 10^5$)—a ground state was found much less frequently.

We have also considered the original $\pm J$ realizations, i.e., the case
$\Delta=0$, with the same number $t_{2,f}=3\times10^5$ of second-type
 annealing.
 The distribution shown in Fig.~\ref{fig:hardness_L16}
reveals that
 the realizations are typically 
 easier than the realizations for $\Delta=1$ obtained
with or without SWAP annealing. Thus, physically, the original instances are
easier to solve than the SWAP-annealed ones. The reason is that
the original problem exhibits a finite GS entropy, i.e., an exponential
number of GSs, which makes finding one of them easier.
Also for excited states encountered during
the annealing many spins can be flipped at zero-energy cost. This
makes  it easier to move through the energy landscape. Nevertheless, with
the SWAP algorithm, i.e., annealing  including the SWAP moves,
the GSs are still easier to find, as can be seen from
Fig.~\ref{fig:p0_tmc_delta10}: for just $t_f=2\times10^3$ in about
95\% of all runs a GS is found. Only when switching off
the SWAP move and using just local spin flips as in the second type
of annealing, finding the ground
state for the annealed realizations becomes harder
as compared to the original ones. This shows
that the improvement by the SWAP annealing originates from actually using the
SWAP moves and not only from the modified bond realizations resulting from
the SWAP moves.

\begin{figure}[htb]
\begin{center}
  \includegraphics[width=0.9\columnwidth]{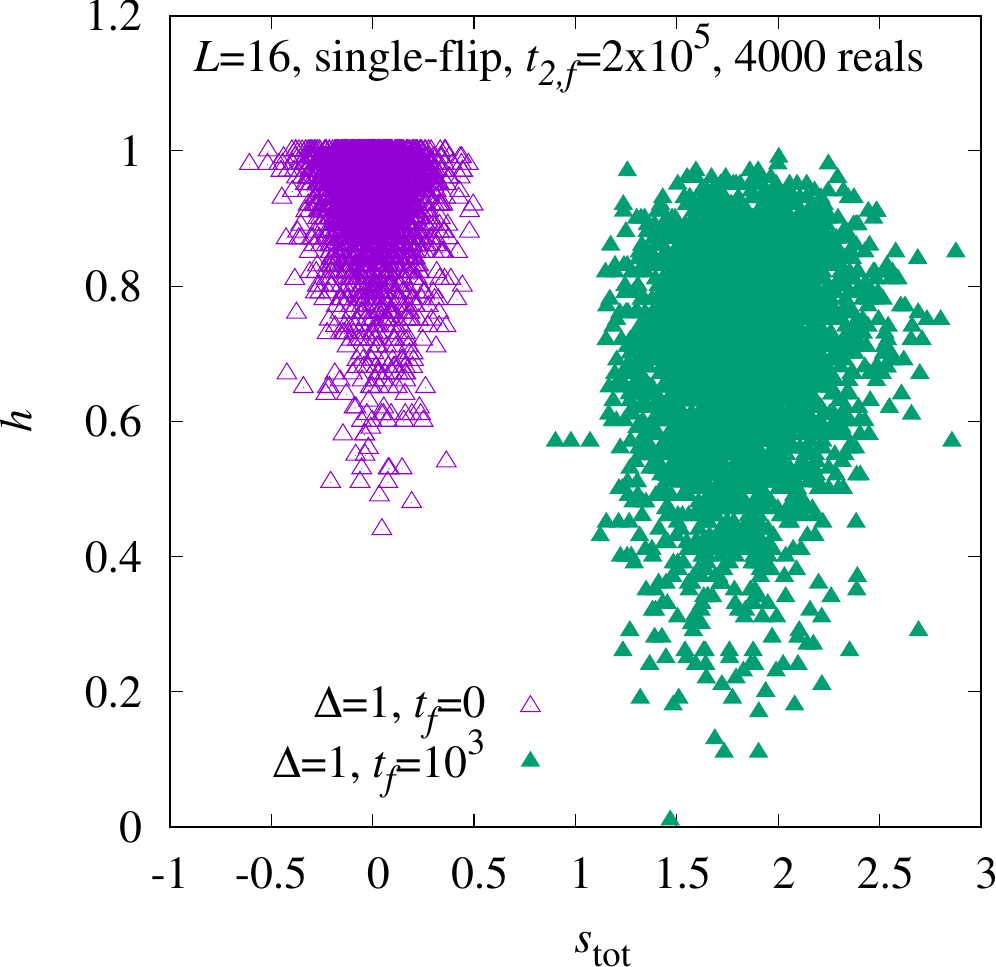}
\end{center}
    \caption{\label{fig:correlation_f_L16}
      The correlations, shown as a scatter plot,
      of the hardness values $h$ for $L=16$ with the total satisfaction
      $s_{\rm tot}$ of each realization.
      Data for the SWAP ensemble
      without annealing ($\Delta=1$, $t_f=0$) and
      the annealed SWAP ensemble with $\Delta=1$
      after $t_f=1000$ annealing
    sweeps.  
    }
\end{figure}

Next, we study whether the hardness is related
to  easy-to-measure parameters
of the realizations. Since the physical behavior is influenced by the
frustration, we first consider the total satisfaction $s_{\rm tot}$, see
Eq.~(\ref{eq:stot:def}). The relation between hardness $h$
and total satisfaction is shown as a scatter plot
in Fig.~\ref{fig:correlation_f_L16} for the 4000
realizations which were obtained
for $L=16$ and $\Delta=1$. Without SWAP annealing, $t_f=0$,
the values of $s_{\rm tot}$ are scattered around 0,  while for
$t_f=1000$, the values are scattered near $s_{\rm tot}=2$, as expected
from the development of $s_{\rm tot}$ with the SWAP annealing time
shown in Fig.~\ref{fig:ftot_tmc_delta10}. The scatter of the hardness
values $h$ seen here
represents the distributions $P(h)$
  shown in Fig.~\ref{fig:hardness_L16}. Thus, for the case $t_f=0$ larger values of $h$ arise as compared to $t_f=1000$. Still, between  $s_{\rm tot}$
  and $h$ no strong correlation is visible within each set
  of realizations of a given annealing schedule. 
  Also, we verified that the measured linear
  correlations are near zero within each set. 

  We obtained a similar result for the original
  $\pm J$ realization, i.e. $\Delta=0$, which is not shown here.
  When considering instead of $s_{\rm tot}$ the
  number of frustrated plaquettes, which does not change by the SWAP
  annealing, or when considering the GS energy density $e_0$
  of each realization,
  we did not observe any correlation with the hardness. Thus,
  the hardness of a realization is a rather complex property of the
  full energy landscape, which is not easily measured without actually
  performing annealing runs.

\begin{figure}[htb]
\begin{center}
  \includegraphics[width=0.9\columnwidth]{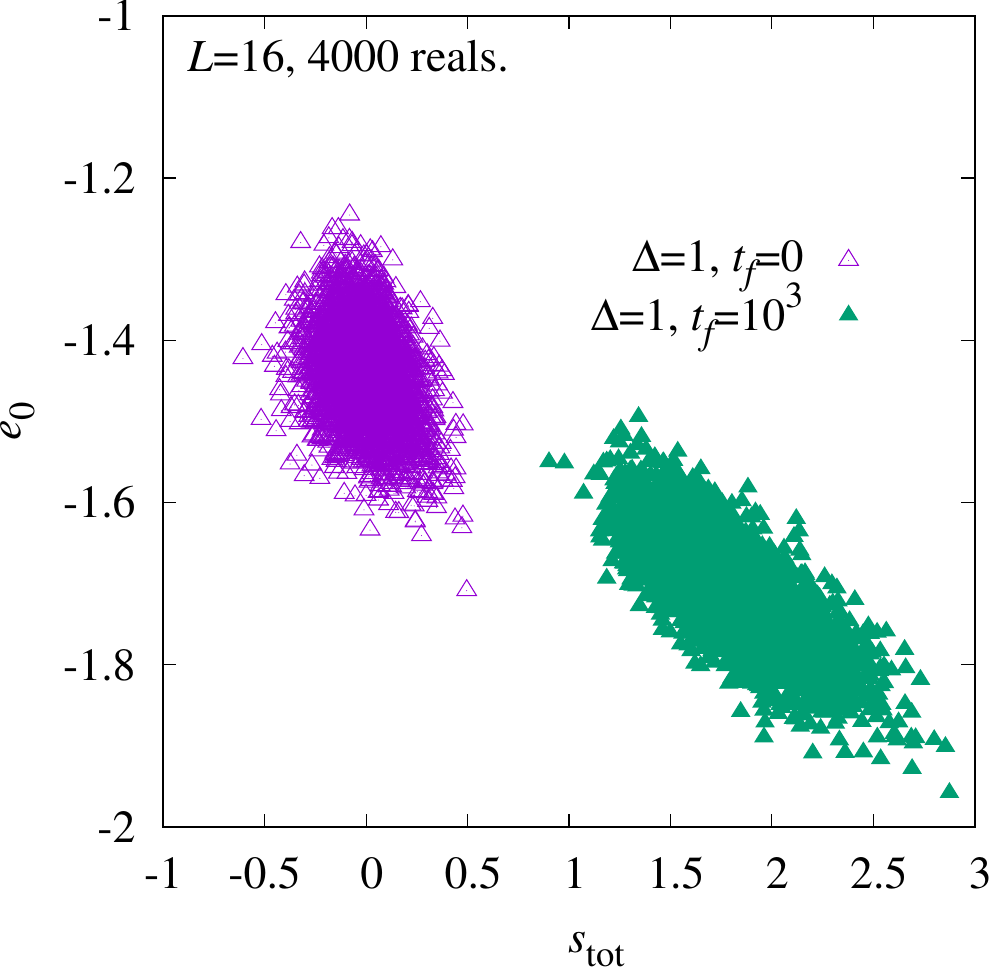}
\end{center}
    \caption{\label{fig:correlation_ef_L16}
      The correlations, shown as a scatter plot,
      of the GS energy density $e_0$
       with the total satisfaction
      $s_{\rm tot}$ of each realization, in systems with $L=16$.
      Shown are  the SWAP ensemble
      without annealing ($\Delta=1$, $t_f=0$) and
      the annealed SWAP ensemble with $\Delta=1$
      after $t_f=1000$ annealing sweeps. 
      }
\end{figure}

  Note, finally, that the total satisfaction actually
  has some relation with properties of the realization, as visible for the GS
  energy density $e_0$, in Fig.~\ref{fig:correlation_ef_L16}.
  Interestingly, the correlation between $e_0$ and $s_{\rm tot}$
  becomes stronger through the SWAP annealing.
  For clarity of the plot the results for $\Delta=0$ are not included
  here.

\subsection{Other Values of $\Delta$}

\begin{figure}[htb]
\includegraphics[width=0.9\columnwidth]{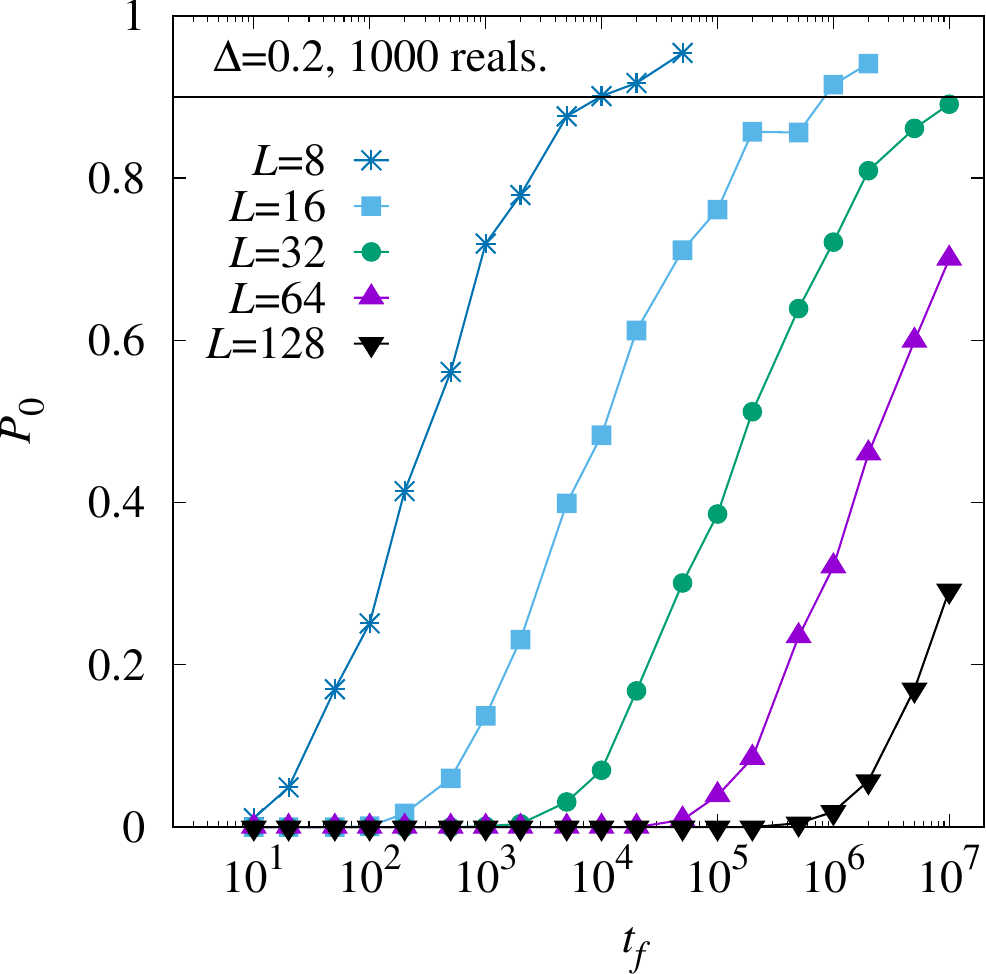}
\caption{\label{fig:p0_tmc_delta02}
  The probability $P_{0}$ that during the SWAP annealing a GS
  is obtained for the resulting disorder realization ${\mathcal J}^*$
  as a function of the number $t_f$ of annealing steps, for
  $\Delta=0.2$ and  different system sizes $L$. Lines are guides to the eyes
only. The horizontal line indicates the 0.95 level for the probability.
}
\end{figure}

As we have seen, the original
$\pm J$ ensemble with $\Delta=0$ behaves very differently
than the SWAP ensemble with $\Delta=1$. On the other hand
for non-zero but small values of $\Delta$, the SWAP moves are less
effective, so
it can
be anticipated that the behavior for small values of $\Delta$
 resembles the $\Delta=0$ case.
Here, we consider the case $\Delta=0.2$.

In Fig.~\ref{fig:p0_tmc_delta02}, the
corresponding results for the probability $P_0$ to find a GS
by the SWAP annealing is shown.
One observes that the annealing time needed to reach 
a GS for many realizations increases considerably as compared to the case of $\Delta=1$,
see Fig.~\ref{fig:p0_tmc_delta10}. To avoid using extreme computational
resources,  we do not perform  SWAP
annealing  beyond $t_f=10^7$ sweeps. 
As visible in the figure, for the largest size $L=128$ considered here, this yields a GS of the annealed realizations ${\cal J}^*$ 
only in $P_0=30\%$ of all cases.
The fact that finding GSs becomes
more difficult can be expected since for small value of $\Delta$ the
influence of the SWAP operation is weak while still the GS degeneracy is lifted. 
A high GS degeneracy, as present for the case $\Delta=0$, would also accelerate determining GSs.

\begin{figure}[ht]
\includegraphics[width=0.9\columnwidth]{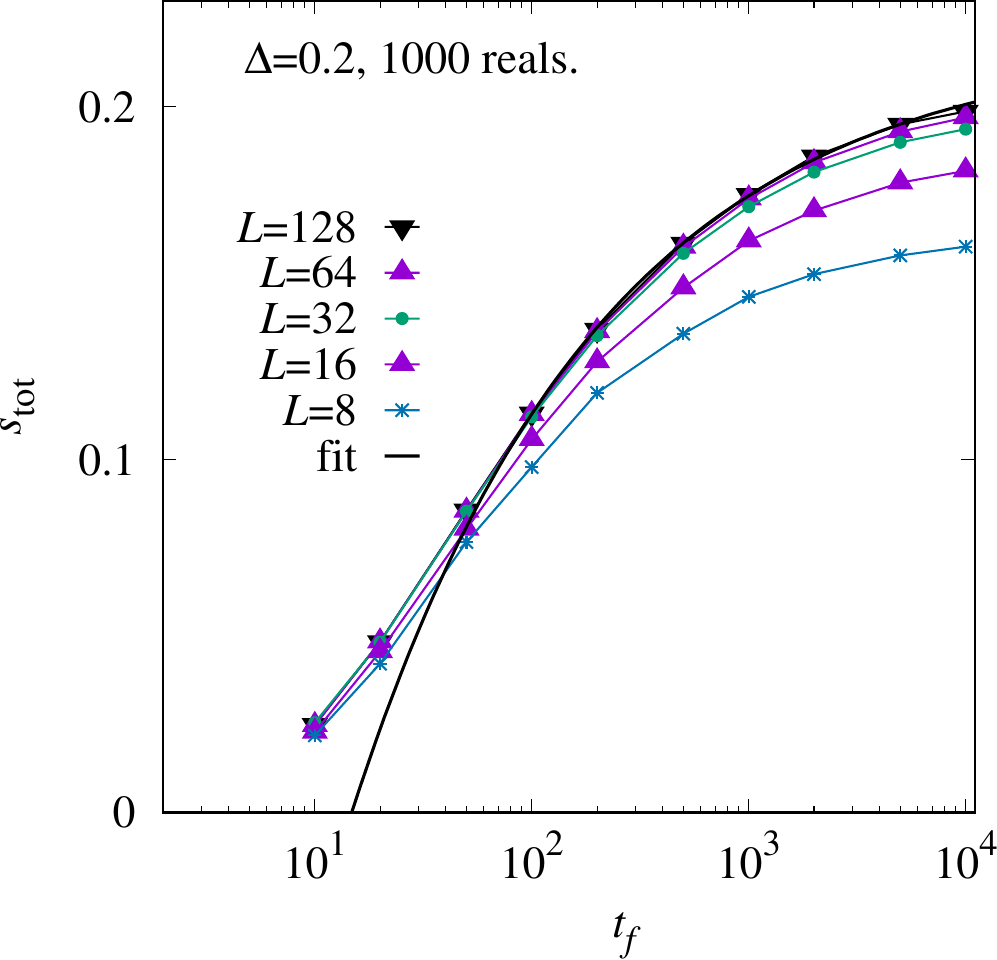}
\caption{\label{fig:ftot_tmc_delta02}
  The average total satisfaction $s_{\rm tot}$ as a function of
  the number $t_f$ of SWAP annealing steps, for
  $\Delta=0.2$ and  different system sizes $L$.
Lines are guides to the eyes only,
      except for the black curve which represents
      a fit of the $L=128$ data to the power-law in Eq.~(\ref{eq:stot:fit}). 
      }
\end{figure}

To understand this behavior, we present
in Fig.~\ref{fig:ftot_tmc_delta02} the satisfaction 
$s_{\rm tot}$ for
the case of $\Delta=0.2$. The curves look very similar
to the ones for the case $\Delta=1$, see Fig.~\ref{fig:ftot_tmc_delta10},
but only much lower levels of $s_{\rm tot}$ are
reached. For $L=128$ a fit to 
power-law Eq.~(\ref{eq:stot:fit}) yields the limiting value
$s_{\rm tot}^{\infty}=0.219(3)$ (with $b=-0.61(4)$ and $c=0.38(2)$), 
which is much lower than the
value $s_{\rm tot}^{\infty}=2.13(1)$ found for
$\Delta=1$. Thus, in the case $\Delta = 0.2$, 
the effective frustration can be reduced only slightly
by the SWAP annealing.

As comparison to what the levels of frustration
might imply, one can consider the case of a $\pm J$ random bond system
on a square lattice 
where a fraction $q$ of all bonds are antiferromagnetic, thus a fraction
of $1-q$ bonds is ferromagnetic. Plaquettes
with one and three antiferromagnetic bonds
are frustrated, which happens with a probability of $4q(1-q)^3$ and
$4q^3(1-q)$, respectively. They contribute a frustration value $f_P=-1$.
All other plaquettes, with zero, two and four antiferromagnetic bonds,
which appear with probabilities $q^4$, $6q^2(1-q)^2$ and $(1-q)^4$,
respectively, contribute a frustration value $f_P=1$, Thus, the expected
satisfaction is $s_{\rm tot}(q)= [q^4+6q^2(1-q)^2+(1-q)^4]$
$-[4q(1-q)^4 +4q^3(1-q)]$, which is one for $q=0$, the ferromagnet,
as well as for $q=1$, the antiferromagnet.  The expected satisfaction takes its minimum value 
at $q=0.5$, where $s_{\rm tot}(0.5)=0$.
For the two-dimensional spin glass model, a transition between ferromagnetic and
$T=0$ spin-glass behavior was found \cite{rand_bond2004} 
by exact GS calculations at a critical
concentration $q_{\rm c}=0.103(1)$. Here, the total satisfaction is
$s_{\rm tot}(0.103)=0.397$, which is rather non-frustrated and well above
the limiting value of $s_{\rm tot}=0.219$ found for $\Delta=0.2$.
On the basis of this comparison, one could
expect a $T=0$ spin-glass behavior for $\Delta=0.2$.

\begin{figure}[ht]
\vspace{0.5cm}
    \includegraphics[width=0.9\columnwidth]{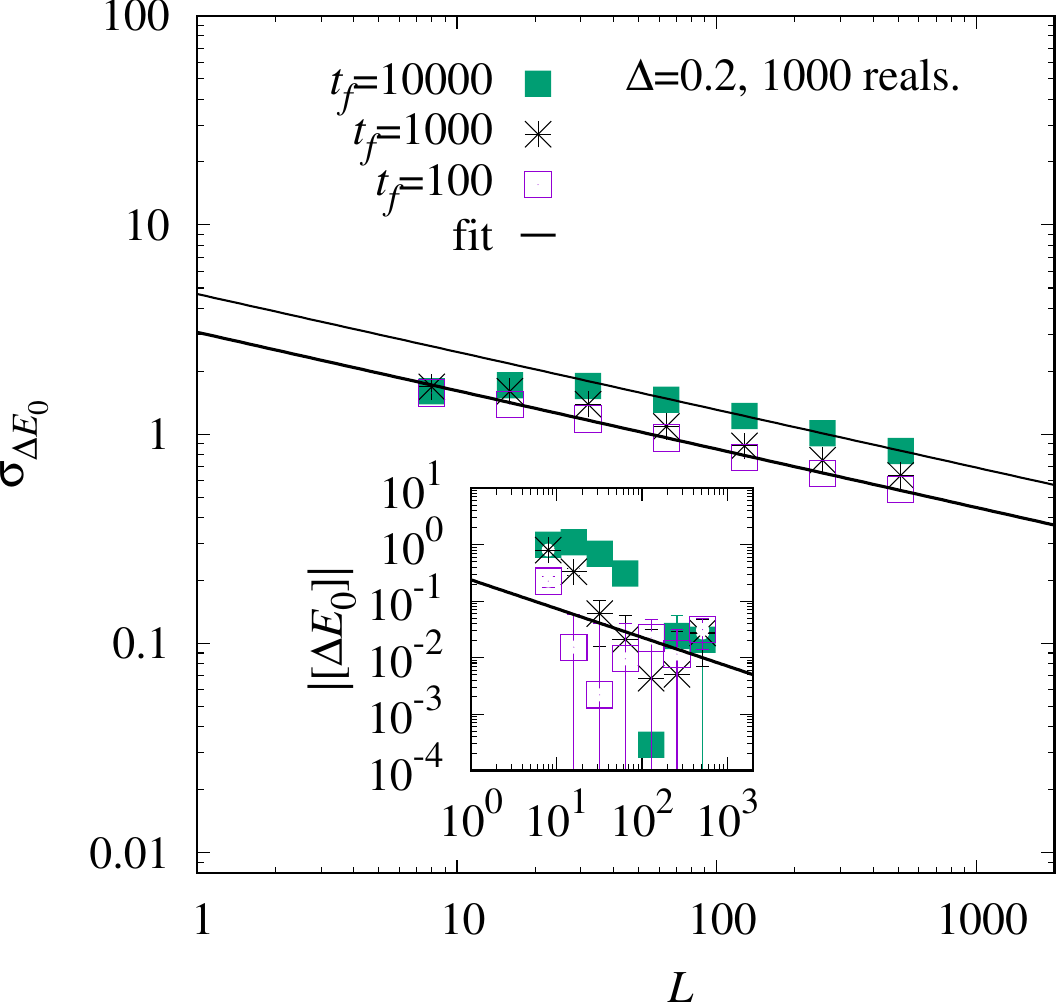}
    \caption{\label{fig:de0_delta02_L_tmc}
      The width $\sigma_{\Delta E_0}$
      of the domain-wall energy distribution as a function of the system
      size $L$, for $\Delta=0.2$
      and different annealing times $t_f$. The inset shows the mean
      $|[ \Delta E_0 ]|$ of
      the distribution again as a function of the system size $L$.
        }
\end{figure}

To investigate whether the SWAP-annealed instances
remain in the $T=0$ spin-glass phase, we studied
the width $\sigma_{\Delta E_0}$
of the domain-wall energy distribution as a function of the system size $L$,
for different annealing times,
see Fig.~\ref{fig:de0_delta02_L_tmc}. In the inset the
average of this distribution is displayed. Even for long annealing
times $t_f=10^4$, the mean and the width decrease with
the system size $L$. The mean is basically close to zero and fluctuates
strongly. Just for completeness, we  have fitted a power law
as Eq.~(\ref{eq:mean:E0:scale}), which resulted in $\tilde \theta=-0.5$
with an error bar of the same size.
More interesting is to fit the power law Eq.~(\ref{eq:def:theta}) to the
variance. For a short SWAP annealing $t_f=100$ we have
found $\theta=-0.279(7)$ for $L\ge 32$,
while for $t_f=100$ we have found $\theta=-0.276(2)$
for $L\ge 128$,
which are both compatible  with the known values
\cite{stiff2d,aspect-ratio2002,khoshbakht2018}
for the Gaussian spin glass.

Thus, for small values of $\Delta$, the system does not loose its
$T=0$ spin-glass
order, even for large annealing times.
This is reflected by the fact that finding GSs using the SWAP algorithm
takes very long, much longer than for the case $\Delta=1$, as discussed
above. 

\begin{figure}[htb]
\vspace{0.5cm}
\includegraphics[width=0.9\columnwidth]{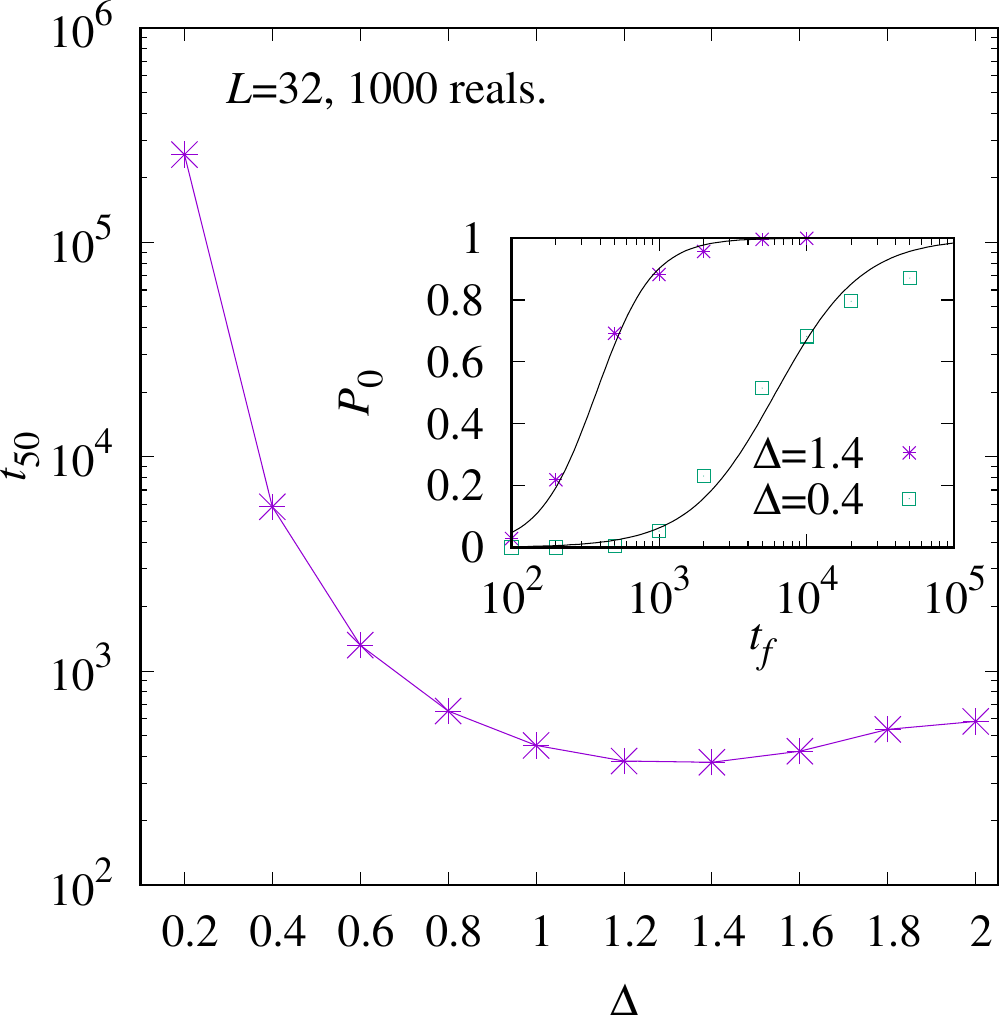}
\caption{\label{fig:t50_L32_delta}
  The annealing time $t_{50}$ required to obtain a GS for 50 \% of the
  realizations of a system with linear size $L=32$ 
  using the SWAP annealing procedure, as a function of
  the length scale $\Delta$. Lines in the main plot are guides to the eyes only.
  The inset shows the probability $P_0$ to obtain a ground state by SWAP
  annealing as a function of the number $t_f$ of annealing sweeps,
  for two values of $\Delta$.
  The lines show fits to Eq.~(\ref{eq:sigmoid}), used to
  determine $t_{50}$.}
\end{figure}

To study how the behavior depends on $\Delta$,
we investigate
how often a GS is found as a function of $\Delta$. In
order to limit the computational effort, we restrict ourselves to
the moderate size $L=32$. This is also sufficient to obtain
qualitative results.
Again, we measure the $P_0$ of finding a GS as a function
of the number of annealing sweeps $t_f$, as shown for
the  examples $\Delta=0.4$ and $\Delta=1.4$
in the inset of Fig.~\ref{fig:t50_L32_delta}.
For each
probability function $P_0(t_f)$, we fit a sigmoidal function
\begin{equation}
  s(t_f)= \frac 1 2 (1+{\rm tanh}( \lambda (t_f-t_{50})))
  \label{eq:sigmoid}
\end{equation}
which determines the typical time scale $t_{50}$ where for about 50 \% of
the realizations
a GS is found by the SWAP annealing. As visible in the inset,
actual data approximately follow the sigmoidal function but not precisely.
Since we are only interested in obtaining a time scale $t_{50}$
in a consistent way, this is sufficient.

The result for $t_{50}$
is shown  in  Fig.~\ref{fig:t50_L32_delta} for $L=32$.
Thus, the time to find a GS
changes only weakly for values of $\Delta$ between
  about 0.8 and 2. On the other hand, for small values of $\Delta$ it increases
  considerably. This goes hand-in-hand with the fact that for small values
  of $\Delta$ the system remains a $T=0$ spin-glass even after long annealing
  with the SWAP algorithm. It is likely that
  there is some ``critical value'' $\Delta_{\rm c}$ which separates
  the two classes. Determining such a critical value would require
  a huge numerical effort and, in our opinion, does not lead to
  more physical insight. Therefore, we did not perform this analysis here.
  
\section{Mean-field estimation}

We have seen that since the early stages of the SWAP dynamics, the system reduces local frustration by rearranging the $\tau_i$ degrees of freedom. Although the total fractions of frustrated ($f_P < 0$) and unfrustrated ($f_P > 0$) plaquettes are each fixed to $1/2$, the SWAP dynamics cause $f_P$ to become less negative on frustrated plaquettes and more positive on unfrustrated ones. As a result, the SWAP dynamics deform the original probability distribution of $f_P$ toward larger values (see Fig.~\ref{fig:frust_distr_delta10}), thereby reducing the total  frustration. When $\Delta$ is sufficiently large, the frustration is so strongly suppressed that the system becomes ferromagnetic and the ground state loses its spin-glass character. The numerical results presented in the previous section clearly indicate a phase transition, 
driven by the effective increase of the satisfaction $s_{\rm tot}$, from the spin-glass behavior (or its two-dimensional vestige) of the original model with quenched $\tau_i$'s toward a conventional ferromagnetic phase, as either the annealing time or $\Delta$ is increased. 

However, locating this transition precisely through numerical simulations alone would be extremely costly. We therefore perform in this section a simple mean-field analysis to estimate the critical value of $s_{\rm tot}$ at which the transition occurs.

To this end, we introduce the following disordered spin model:
\begin{equation} \label{eq:HMF}
    {\cal H}_{\rm est} = - \sum_{\langle ij \rangle} J_{ij} \, s_i s_j \, ,
\end{equation}
where the couplings $J_{ij}$ are drawn independently from the bimodal distribution:
\begin{equation}
    p(J_{ij}) = \frac{1}{2} \delta (J_{ij} - J) + \frac{1}{2} \delta (J_{ij} + \phi J) \, ,
\end{equation}
with $\phi \in [0,1]$ a tunable asymmetry parameter. For $\phi = 1$ one recovers the standard EA model with $\pm J$ couplings, while $\phi = 0$ reduces to a diluted ferromagnet with half the bonds removed. This choice is motivated by the desire to mimic the asymmetric shape of the distribution shown in Fig.~\ref{fig:frust_distr_delta10}, while keeping the fractions of frustrated and unfrustrated plaquettes fixed to $1/2$. The induced probability distribution of the plaquette variable $f_P$ reads
\begin{equation}
\begin{aligned}
    p(f_P) =\;
         & \frac{1}{16}\,\delta(f_P - J^4) 
             + \frac{1}{4}\,\delta(f_P + \phi J^4) 
            \\
            & 
             + \frac{3}{8}\,\delta(f_P - \phi^2 J^4) + \frac{1}{4}\,\delta(f_P + \phi^3 J^4) 
             \\
            &
             + \frac{1}{16}\,\delta(f_P - \phi^4 J^4) \, ,
\end{aligned}
\label{eq:pfP}
\end{equation}
from which one obtains
\begin{equation}
    s_{\rm tot} = [\langle f_P \rangle_P] = \frac{J^4}{16} \left(1 - \phi\right)^4 \, .
\label{eq:fPavg}
\end{equation}

\begin{figure}[htb]
\vspace{0.5cm}
\includegraphics[width=0.9\columnwidth]{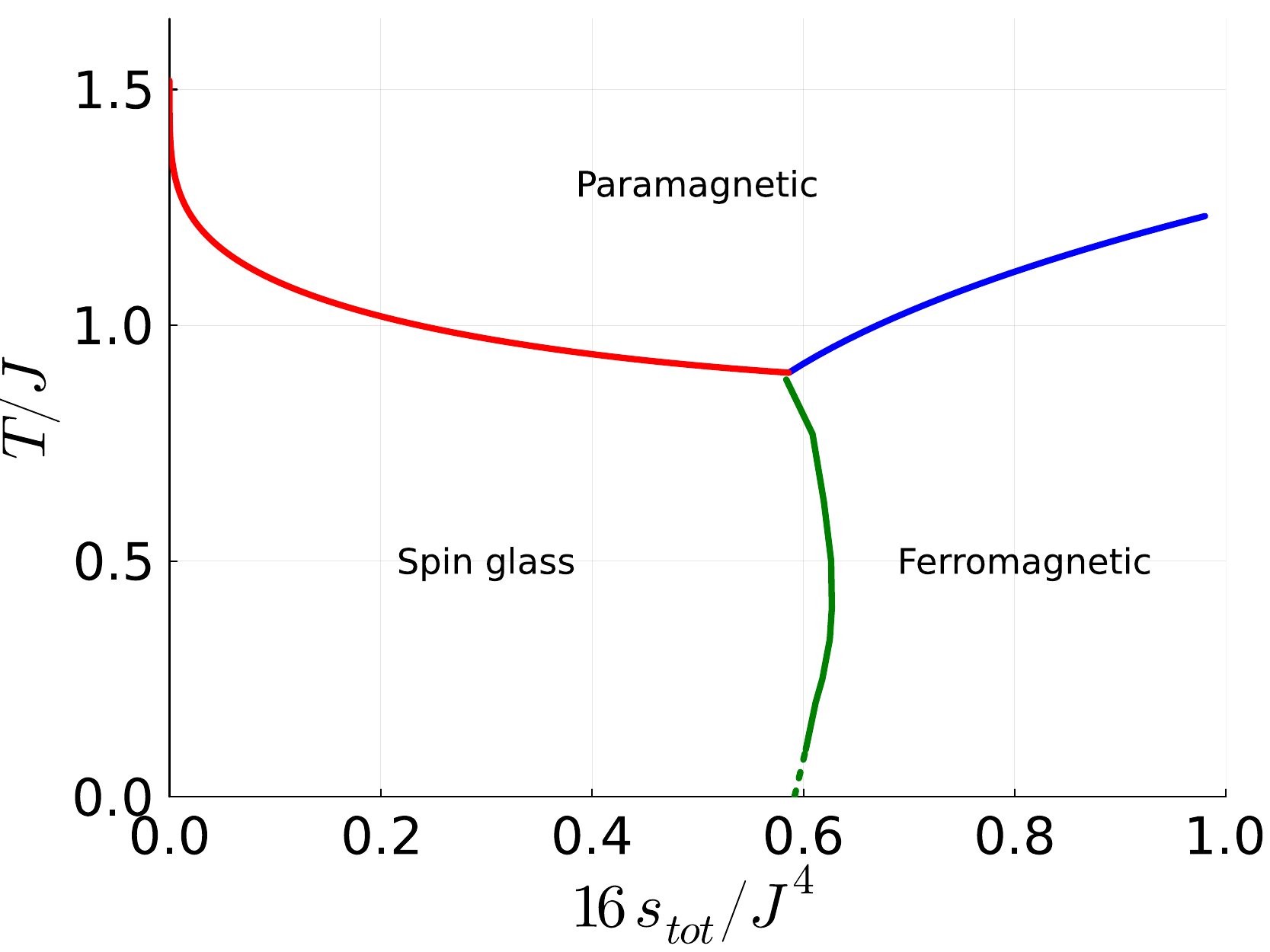}
\caption{\label{fig:mean_field_phase_diagram}
Mean-field phase diagram of the model~\eqref{eq:HMF} on the Bethe lattice with branching ratio $k = 3$, displayed in the $16 s_{\rm tot}/J^4$-$T/J$ plane.
The low $T$ part of the green line is dotted purely as a guide to the eye, as the population dynamics algorithm suffers from poor numerical convergence at very low temperatures.}
\end{figure}

The simplest mean-field treatment consists in solving this model on the Bethe lattice. To mimic the $2D$ Euclidean lattice, we fix the connectivity to $k+1 = 4$, corresponding to a branching ratio $k = 3$. On the Bethe lattice, the solution is encoded in a set of self-consistent equations for the cavity magnetizations $m_{i \to j} = \langle s_i \rangle_{\setminus j}$, defined as the local magnetization on site $i$ in the absence of its neighbor $j$. We refer the reader to Refs.~\cite{mezard2001bethe, mezard1987} for a detailed derivation, and only quote here the main result. The cavity equation reads:
\begin{equation} \label{eq:cavity}
    m_{i \to j} = \tanh \!\left( \sum_{\ell \in \partial i \setminus j} 
    \mathrm{atanh} \!\left( \tanh\!\left(\beta J_{\ell i}\right) m_{\ell \to i} \right) \right) .
\end{equation}
The trivial fixed point $m_{i \to j} = 0$ always exists and corresponds to the paramagnetic phase. The ferromagnetic instability is signaled by the growth of the average $\langle m_{i \to j} \rangle$ under iteration, while the spin-glass instability corresponds to the growth of its variance while the average remains zero. The two transition lines from the high-temperature paramagnetic phase to the low-temperature ferromagnetic and spin-glass phases respectively are obtained from a linear stability analysis of the first two moments of the cavity magnetization around zero~\cite{mezard2001bethe}, yielding:
\begin{equation}
\begin{aligned}
    & \tanh(\beta_{\rm FM} J) - \tanh(\beta_{\rm FM} \phi J) = \frac{2}{k} \, , \\[4pt]
    & \tanh^2(\beta_{\rm SG} J) + \tanh^2(\beta_{\rm SG} \phi J) = \frac{2}{k} \, .
\end{aligned}
\label{eq:transition_lines}
\end{equation}
The numerical solutions of these equations are shown in the phase diagram of Fig.~\ref{fig:mean_field_phase_diagram} with 
a blue and a red line, respectively.
In the figure, in place of $\phi$, we use $16 s_{\rm tot}/J^4$ as the control parameter via Eq.~\eqref{eq:fPavg}, so as to make direct contact with the numerical results of the previous sections.

There exists also a transition line separating the ferromagnetic phase from the spin-glass phase at low temperatures. Obtaining this transition line is slightly more involved, as one must solve Eq.~\eqref{eq:cavity} when $m_{i \to j} \neq 0$ and the cavity fields vary from node to node. This transition line cannot therefore be obtained from a linear stability analysis around the paramagnetic fixed point; instead, the cavity equation must be solved explicitly using the population dynamics algorithm~\cite{mezard2001bethe}. The transition from the ferromagnetic phase to the spin-glass phase is then identified either by the vanishing of the average magnetization or by the divergence of the spin-glass susceptibility~\cite{mezard2001bethe, mezard1987}. The results of this procedure are also shown in Fig.~\ref{fig:mean_field_phase_diagram} with a green line.

The phase diagram reveals that, upon increasing $s_{\rm tot}$, the spin-glass phase gives way to a conventional ferromagnetic phase. In particular, the ground state undergoes a spin-glass-to-ferromagnet transition at $T = 0$ when $16 s_{\rm tot}/J^4 \sim 0.6$.
For $J=1$ the mean-field 
critical value is therefore $s_{\rm tot} \sim 0.0375$. In the SWAP simulations for the two-dimensional lattice we still find 
spin-glass behavior for a much larger value $s_{\rm tot}= \left[ \left\langle f_{P}
    \right\rangle_{P} \right]\approx 0.2$ (for $\Delta=0.2$), whereas ferromagnetic behavior for $s_{\rm tot}\approx 2.1$ (for $\Delta=1$).
   Quantitative agreement between the mean-field calculation and the finite-dimensional lattice simulations is not expected, in particular because the probability distribution $f(f_P)$ of the toy model, Eq.~\eqref{eq:pfP}, differs significantly from those shown in Fig.~\ref{fig:frust_distr_delta10}. Nevertheless, the two approaches consistently indicate the existence of such a transition.

\section{Summary and Discussion\label{sec:discussion}}

We have considered 
the SWAP ensemble of spin glasses, which was recently introduced~\cite{miranda2024}, 
focusing on the two-dimensional square lattice case.
Our three main questions were: does the SWAP algorithm
obtain ground states even for large system sizes,
how long does it take, and what are the
physical properties of the 
annealed bond realizations obtained in this manner?

By comparing with exact ground states obtained from a polynomial-time
minimum-weight perfect matching algorithm, we showed that the
SWAP algorithm is indeed able to obtain the ground states of large Ising systems
with the bond configurations ${\mathcal J}^*$ found after annealing.
However, it does at the cost of an annealing time
which increases exponentially with system size. Thus, SWAP annealing
does not make the GS problem ``easy'', i.e. polynomial-time,
when one uses 
SWAP and physical spin-flip updates. This is in contrast to the existence
of a  polynomial-time exact ground state algorithm.

To investigate the physical behavior of the resulting Ising models with the 
bond realizations ${\mathcal J}^*$  obtained after the SWAP annealing, we studied 
the domain-wall properties by using exact
ground-state calculations in combination with
switching the boundary conditions from periodic to anti-periodic.
Since the matching algorithm is fast, large system sizes such as
$L=1024$ could be treated.

For large enough values of the
spin length scale $\Delta$, here we studied $\Delta=1$,
the mean of the domain-wall energy $\Delta E_0$ increases
linearly with system size, i.e. the realizations behave like ferromagnets,
actually gauge-transformed ferromagnets since half of the
bonds have a negative sign. 
Thus, the SWAP annealing destroys the $T=0$ spin-glass behavior. This happens
although half of the plaquettes are still frustrated.

On the other hand, for short annealing times or small values 
of the spin length scale $\Delta$, the realizations ${\mathcal J}^*$ still behave
like $T=0$ spin glasses.
For these parameters, the SWAP annealing time needed to find ground states also
increases exponentially, and is even longer than for $\Delta =1$.
Thus, we expect a critical value $\Delta_{\rm c}$ below which the
$T=0$ spin-glass behavior  remains, at the cost of very long annealing times needed.

We have computed GSs for the annealed bonds ${\mathcal J}^*$ 
which are fixed as obtained after the SWAP annealing with only the Ising variables $\sigma$ being the degrees of freedom. One could in principle try to answer the question about the nature of the low-energy states in the same fashion by GS calculations of the original Hamiltonian Eq.~(\ref{eq:H}) with both the length variables $\tau$ and the Ising variables $\sigma$ subject to optimization. Since the length variables are not totally free to adjust, they can only be exchanged, this is not trivial to achieve. We are not aware of an efficient algorithm to obtain these GSs. Likely this GS problem is NP-hard \cite{garey1979}, i.e., only exponential-time algorithms are known. 
Nevertheless, because the ground states (GSs) obtained for the Hamiltonian in
Eq.~(\ref{eq:H:annealed})
optimize over a smaller set of variables, their calculated ground-state energies $E_0$ serve as an upper bound for the true ground-state energies of the original Hamiltonian.
Actually, if the number of annealing steps is large, leading to low-energy choices for the length variables $\tau$, the bound will be very close to the actual value. Now, with large-enough value of $\Delta$ and large-enough number of annealing sweeps $t_f$ the SWAP ensemble behaves as a gauge-transformed ferromagnetic one. Because lowering the energy even more will make it even more ferromagnetic, see Fig.~\ref{fig:corr_e0_de0_delta10}, we expect that also the original Hamiltonian with both $\tau$ and $\sigma$ as degrees of freedom shows ferromagnetic behavior.

The primary reason for using SWAP methods is to efficiently sample supercooled liquid or glassy configurations while bypassing the dramatic slowdown of physical glass dynamics~\cite{berthier2016,ninarello2017,berthier_efficient_2019,Scalliet22}. In this paper, we showed that introducing a continuous polydispersity in spin length and enabling non-local SWAP moves---in a manner that mimics the strategy employed for particle glass formers—retains the spin---glass nature of the ground states only for weak spin-length polydispersity. For stronger polydispersity, however, frustration is spatially reorganized during the SWAP evolution, ultimately driving the system toward ferromagnetic ground states. One might wonder whether a similar phenomenon occurs in particle glass formers---namely, whether increasing the polydispersity beyond a certain threshold causes the system to lose its glassy properties. However, key observables used here to characterize ground states, such as local frustration and the domain-wall energy distribution, lack a direct analogue in particle systems. Yet, it would be extremely interesting to systematically investigate whether SWAP dynamics in polydisperse particle glass formers induces spatial correlations in particle sizes---a feature that would serve as the signature of the formation of small, highly unfrustrated domains, similar to those shown in Fig.~\ref{fig:frustration_pattern}.

It was argued in~\cite{Ghimenti26} that the SWAP algorithms applied to supercooled liquids and glasses
act by exploiting the time reparametrization softness of these systems~\cite{Chamon02,Chamon04,Castillo07,Chamon07,Dyre22}. 
In so doing, they 
would change the speed at which trajectories in configuration space are followed as the system slowly relaxes. 
We found no evidence for this mechanism in the SWAP for disordered spin model that we implemented
here.

In the future,  it would be of interest to study spin-glasses in higher
dimensions like $d=3$. This is  more demanding numerically, because
no polynomial-time ground state algorithm is available. Thus, one
can address only small system sizes, and it is questionable whether
one can reach the limiting behavior of the domain-wall energy distribution,
which is needed to distinguish a ferromagnet from a spin-glass.

Furthermore, we plan to implement a parallel tempering approach, where
different values of $\Delta$ are considered simultaneously, for a given
realization of the bonds, to investigate whether this helps
in obtaining low-energy states for models close to the original one, 
especially for $\Delta=0$.

\begin{acknowledgments}

  AKH thanks the \emph{Laboratoire de Physique Théorqiue de la
  Matièere Condensée} (LPTMC)
  and the \emph{Laboratoire de Physique Théorqiue des Hautes Energies} (LPTHE)
  at the University Sorbonne, Paris, where
  the main parts of the work were performed, for their hospitality
  and the financial support. The work was also supported by
  the \emph{Centre National de Recherche Scientifique}, France (CNRS).
    The simulations were performed at the
    HPC cluster ROSA, located at the University of Oldenburg (Germany) and
    funded by the DFG through its Major Research Instrumentation Program
    (INST 184/225-1 FUGG) and the Ministry of Science and Culture (MWK) of the
    Lower Saxony State. LFC and MT acknowledge funding from the ANR research 
    grant Many-BodyNet ANR-24-CE30-5851. LFC and MT thank Greivin Alfaro Miranda for previous 
collaboration on this subject. 
\end{acknowledgments}

\vspace*{3mm}
\centerline{\bf DATA AVAILABILITY}

\vspace*{2mm}

The gnuplot plot files and the data generated in the simulations which
are used in the plot files are publicly available in the DARE
repository of the University of Oldenburg \cite{data_sg_swap}

\bibliography{refs.bib}

\end{document}